\documentclass[a4paper,final]{article}
\usepackage{amsfonts,amsmath,amsthm,amssymb,xcolor,colortbl}
\usepackage{enumitem}
\usepackage{calrsfs}
\DeclareMathAlphabet{\pazocal}{OMS}{zplm}{m}{n}

\usepackage{geometry}
\usepackage[onehalfspacing]{setspace}
\usepackage{soul}
\usepackage{setspace}
\usepackage{natbib}
\usepackage{clipboard}	
\usepackage{graphicx}
\newclipboard{PaperClipboard}	
\usepackage{xr-hyper} 
\usepackage[colorlinks,linkcolor=links,citecolor=cites,urlcolor=MyDarkBlue]{hyperref}
\usepackage{cleveref}
\usepackage{bbm}
\usepackage[justification=centering]{caption}
\usepackage[justification=centering]{subfig}
\usepackage[section]{placeins}
\usepackage{parskip}
\usepackage{nicefrac}
\usepackage{appendix}
\usepackage[multiple]{footmisc}
\definecolor{MyDarkBlue}{rgb}{0,0.08,0.45}
\definecolor{cites}{HTML}{324b13}
\definecolor{links}{HTML}{1a663b}
\definecolor{MyLightMagenta}{cmyk}{0.1,0.8,0,0.1}
\definecolor{sblue}{HTML}{0049A9}
\definecolor{scyan}{HTML}{CBEAFC}
\definecolor{sred}{HTML}{B5595C}
\definecolor{sgreen}{HTML}{609B57}
\definecolor{spink}{HTML}{FFB0FF}
\newtheorem{theorem}{Theorem}
 
\newtheorem{proposition}{Proposition}
\newtheorem{example}{Example}
\newtheorem{definition}{Definition} 
\newtheorem{corollary}{Corollary}

\newtheorem{remark}{Remark}
\newtheorem{observation}{Observation}

\newcommand{\genericu}{\ensuremath{u(a,b)}}

\newcommand{\genericv}{\ensuremath{v(a,b)}}

\newcommand{\cala}{\ensuremath{\pazocal{A}}}
\newcommand{\calb}{\ensuremath{\pazocal{B}}}

\newcommand{\sidea}{\ensuremath{\mathcal{A}}}
\newcommand{\sideb}{\ensuremath{\mathcal{B}}}

\newcommand{\arrivalaupto}{\ensuremath{\overline{A}}}
\newcommand{\arrivalbupto}{\ensuremath{\overline{B}}}
\newcommand{\arrivalauptot}{\ensuremath{\arrivalaupto_t}}
\newcommand{\arrivalbuptot}{\ensuremath{\arrivalbupto_t}}

\newcommand{\matching}{\ensuremath{m}}
\newcommand{\matchingb}{\ensuremath{\overline{\matching}}}
\newcommand{\matchingbt}{\ensuremath{\matchingb_t}}
\newcommand{\matchingbuptot}{\ensuremath{\matchingb^t}}
\newcommand{\matchingc}{\ensuremath{\widetilde{\matching}}}
\newcommand{\matchingstar}{\ensuremath{\matching^\star}}

\newcommand{\matchingt}{\ensuremath{\matching_t}}

\newcommand{\matchinguptot}{\ensuremath{\matching^{t-1}}}

\newcommand{\terminal}{\ensuremath{T}}
\newcommand{\terminalb}{\ensuremath{\terminal^\prime}}

\newcommand{\Matchings}{\ensuremath{\pazocal{M}}}
\newcommand{\Matchingsfinal}{\ensuremath{\Matchings_\terminal}}

\newcommand{\periodone}{\ensuremath{t=1}}
\newcommand{\periodtwo}{\ensuremath{t=2}}

\newcommand{\dateindex}{\ensuremath{s}}

\newcommand{\agent}{\ensuremath{k}}

\newcommand{\agenta}{\ensuremath{a}}

\newcommand{\agentb}{\ensuremath{b}}

\newcommand{\economyr}{\ensuremath{E}}
\newcommand{\economyrtwo}{\ensuremath{\economyr^2}}

\newcommand{\economyrterminal}{\ensuremath{\economyr^\terminal}}

\newcommand{\arrivala}{\ensuremath{A}}
\newcommand{\arrivalat}{\ensuremath{\arrivala_t}}
\newcommand{\arrivalb}{\ensuremath{B}}
\newcommand{\arrivalbt}{\ensuremath{\arrivalb_t}}

\newcommand{\coalition}{\ensuremath{C}}
\newcommand{\coalitionb}{\ensuremath{\coalition^\prime}}

\newcommand{\stable}{\ensuremath{\mathrm{S}}}
\newcommand{\addon}{\ensuremath{F}}

\newcommand{\pair}{\ensuremath{(\agenta,\agentb)}}

\newcommand{\payoffa}{\ensuremath{u}}

\newcommand{\payoffact}{\ensuremath{\payoffa_{\conjecturet}}}
\newcommand{\payoffacb}{\ensuremath{\payoffa_{\conjecturebeg}}}

\newcommand{\payoffb}{\ensuremath{v}}

\newcommand{\payoffbct}{\ensuremath{\payoffb_{\conjecturet}}}
\newcommand{\payoffbcb}{\ensuremath{\payoffb_{\conjecturebeg}}}

\newcommand{\naturals}{\ensuremath{\mathbb{N}}}

\newcommand{\sol}{\ensuremath{\Sigma}}
\newcommand{\conjecture}{\ensuremath{\varphi}}
\newcommand{\conjectures}{\ensuremath{\Phi}}

\newcommand{\agentbb}{\ensuremath{\agentb^\prime}}

\newcommand{\solone}{\cvr\ dynamic stability}

\newcommand{\solonec}{\ensuremath{\conjecture_{\dsh}}}
\newcommand{\solonect}{\ensuremath{\conjecture_{\dsh,t}}}
\newcommand{\soltwo}{sophisticated dynamic stability}
\newcommand{\Soltwo}{Sophisticated dynamic stability}
\newcommand{\sds}{\ensuremath{\mathrm{SDS}}}
\newcommand{\dsh}{\ensuremath{\hat{\ds}}}
\newcommand{\sdsc}{\ensuremath{\conjecture_{\sds}}}
\newcommand{\dsc}{\ensuremath{\conjecture_{\ds}}}
\newcommand{\dsct}{\ensuremath{\conjecture_{\ds,t}}}
\newcommand{\ds}{\ensuremath{\mathrm{DS}}}

\newcommand{\re}{\ensuremath{\mathrm{RE}}}

\newcommand{\calat}{\ensuremath{\cala_t}}
\newcommand{\calbt}{\ensuremath{\calb_t}}
\newcommand{\conjectureb}{\ensuremath{\conjecture}}
\newcommand{\conjecturebeg}{\ensuremath{\conjecture_1}}
\newcommand{\conjecturet}{\ensuremath{\conjecture_t}}
\newcommand{\cvr}{\ensuremath{\mathrm{CVR}}}
\newcommand{\agree}{\ensuremath{\conjecture^\dagger}}
\newcommand{\Agree}{\ensuremath{\conjectures_\dagger}}
\newcommand{\solp}{\ensuremath{\sol^{\Agree}}}
\newcommand{\candidate}{\ensuremath{M^\star}}
\newcommand{\matchingstaruptot}{\ensuremath{\matching^{\star^{t-1}}}}

\usepackage{mparhack}

 \usepackage{chngcntr}
\usepackage{apptools}
\AtAppendix{\counterwithin{equation}{section}}
\AtAppendix{\counterwithin{figure}{section}}
\AtAppendix{\counterwithin{table}{section}}
\AtAppendix{\counterwithin{example}{section}}
\AtAppendix{\counterwithin{remark}{section}}
\AtAppendix{\counterwithin{proposition}{section}}
\AtAppendix{\counterwithin{theorem}{section}}
\AtAppendix{\counterwithin{corollary}{section}}
\usepackage{etoolbox}
\makeatletter
\patchcmd{\hyper@makecurrent}{%
    \ifx\Hy@param\Hy@chapterstring
        \let\Hy@param\Hy@chapapp
    \fi
}{%
    \iftoggle{inappendix}{
        \@checkappendixparam{chapter}%
        \@checkappendixparam{section}%
        \@checkappendixparam{subsection}%
        \@checkappendixparam{subsubsection}%
        \@checkappendixparam{paragraph}%
        \@checkappendixparam{subparagraph}%
    }{}%
}{}{\errmessage{failed to patch}}

\newcommand*{\@checkappendixparam}[1]{%
    \def\@checkappendixparamtmp{#1}%
    \ifx\Hy@param\@checkappendixparamtmp
        \let\Hy@param\Hy@appendixstring
    \fi
}
\makeatletter

\newtoggle{inappendix}
\togglefalse{inappendix}

\apptocmd{\appendix}{\toggletrue{inappendix}}{}{\errmessage{failed to patch}}
\apptocmd{\subappendices}{\toggletrue{inappendix}}{}{\errmessage{failed to patch}}

\title{Consistent Conjectures in Dynamic Matching Markets\footnote{The latest version of this paper can be found \href{https://www.dropbox.com/scl/fi/o7sht11dbzvnufrj37ud8/ccdmm-public.pdf?rlkey=zzxete3g4lo3x3tx51dgi28cd&dl=0}{here}. We thank the Editor, Juan Moreno, an Associate Editor, and three referees for feedback that has greatly improved this paper. We also thank Antonio Nicol\`o, Pietro Salmaso, and Riccardo Saulle for their hospitality and insightful conversations that, among other things, shaped the illustrations in the paper. Finally, we thank the Sloan Foundation for financial support.}}
\author{Laura Doval\thanks{Economics Division, Columbia Business School and CEPR. E-mail: \href{mailto:laura.doval@columbia.edu}{\texttt{laura.doval@columbia.edu}}} \and Pablo Schenone\thanks{Department of Economics, Fordham University. E-mail: \href{mailto:pschenone@fordham.edu}{\texttt{pschenone@fordham.edu}}}}

\newcommand{\recbeg}{\ensuremath{\conjecture_{\re,1}}}

\begin{document}
\maketitle
\begin{abstract}
We provide a framework to study stability notions for two-sided dynamic matching markets in which matching is one-to-one and irreversible. The framework gives center stage to the set of matchings an agent anticipates would ensue should they remain unmatched, which we refer to as the agent's \emph{conjectures}. A collection of conjectures, together with a pairwise stability and individual rationality requirement given the conjectures, defines a solution concept for the economy. We identify a sufficient condition---\emph{consistency}---for a family of conjectures to lead to a nonempty solution (cf. \citealp{hafalir2008stability}). As an application, we introduce two families of consistent conjectures and their corresponding solution concepts: continuation-value-respecting dynamic stability, and the extension to dynamic markets of the solution concept in \cite{hafalir2008stability}, \soltwo.
%
%
\end{abstract}

\section{Introduction}\label{sec:intro}
In this paper, we provide a unifying framework to study solution concepts for two-sided, one-to-one, dynamic matching markets in which matching is irreversible. A key ingredient in any such solution concept is what matchings an agent expects would ensue should they remain unmatched and wait for a better matching opportunity; we dub this set of matchings the agent's \emph{conjectures}. Following the literature on matching with externalities \citep{sasaki1996two,hafalir2008stability}, we take conjectures as a primitive and define a solution concept given these conjectures. We provide a sufficient condition these conjectures must satisfy for the solution concept to be nonempty. Armed with this result, we propose two solution concepts for dynamic matching markets with nonempty solutions.

Formally, we study a two-sided, dynamic matching market, in which agents on both sides of the economy arrive over time. We assume arrivals are deterministic; that is, agents can perfectly foresee when each agent arrives in the economy.\footnote{We assume deterministic arrivals because doing so simplifies notation, but this assumption is not needed for our results. See \autoref{remark:ass}.} A matching specifies who matches with whom in each period, subject to two constraints: agents cannot be matched before they arrive, and agents can have at most one matching partner on the opposite side. Agents are discounted utility maximizers, and their payoff from a matching depends only on whom they match with and when. Our goal is to study solution concepts for this economy.


In \autoref{sec:conjectures}, we introduce the two main ingredients of the model: an agent's conjectures (\autoref{definition:conjectures}) and a solution concept given the conjectures (\autoref{definition:sol-given-c}). When determining whether a matching is self-enforcing, and hence a solution for the economy, the matching should at the very least be immune to two kinds of blocks: blocks by individual agents who may not find their current matching partner acceptable, and blocks by pairs of contemporaneous agents who would rather match together. As we explain next, an agent's conjectures allow us to define the agent's set of acceptable matching partners. We say a matching is a solution for the economy given the conjectures if no agent is matched to an unacceptable matching partner, and no two agents who are available to match at the same time prefer to match together. As we explain in \autoref{sec:conjectures}, in one-period economies, the solution concept given the conjectures reduces to the notion of stability in \cite{gale1962college}, independently of the conjectures.

An agent's conjectures describe the set of matchings the agent assumes may ensue should they remain unmatched in a given period. Importantly, a given conjectured matching describes both the current-period's matching outcome \emph{and} the continuation matching from the next period onward. Because the set of agents available to match from the next period onward depends on both the newly arriving agents and those who remain unmatched from previous periods, the current-period's matching and the next period's arrivals define the set of feasible continuation matchings. In other words, an agent's outcome from remaining unmatched cannot be defined independently of other contemporaneous agents' matching outcomes. Thus,  having assumed away payoff externalities, an externality still exists in our model because other contemporaneous agents' matching outcomes determine which continuation matchings are feasible. For a given conjectured matching, we can define an agent's acceptable partners as those preferred to the conjectured matching. Thus, to determine an agent's acceptable partners, we need an assumption on how to select a continuation matching from the agent's conjectures. Following \cite{sasaki1996two}, we say the agent finds a matching partner acceptable if the matching partner is preferred to some conjectured continuation matching.  


The paper's main result, \autoref{theorem:consistent-conjectures}, provides a sufficient condition on the family of conjectures for matchings satisfying the solution concept to exist. We refer to the conjectures satisfying this property as \emph{consistent} (\autoref{definition:consistent-conjectures}).  Consistent conjectures guarantee that \emph{candidate} matchings, which can be computed from the model primitives, are a solution for the economy (see \autoref{sec:consistent-conjectures} for a definition of the set of candidate matchings). Loosely speaking, consistency requires that each candidate matching\textemdash denote it by \matchingstar\textemdash and the conjectures satisfy the following condition.  In each period and for each agent who is unmatched under \matchingstar\ in that period, the candidate matching \matchingstar\ should be one of the agent's conjectures. In other words, when checking the continuation candidate matching against their own conjectures for what matchings would ensue given that they remain unmatched, the agents who are unmatched under \matchingstar\ will find the candidate matching to be \emph{consistent} with their conjectures. Consistency is related to the condition in  \citet[Proposition 2]{hafalir2008stability}, which ensures the existence of stable matchings in static markets with payoff externalities; we explain the connection between the two in \autoref{sec:consistent-conjectures}.


 Whereas consistency is defined relative to the candidate matching, an immediate corollary is that the following conjectures always lead to a nonempty solution concept. Suppose each agent's conjectures only rule out continuation matchings that do not conform with the solution concept in the continuation. For instance, in a two-period economy, these conjectures specify that in the last period, only stable matchings among the remaining unmatched agents and the new arrivals are possible. These conjectures always satisfy consistency, and \autoref{theorem:consistent-conjectures} implies they lead to a nonempty solution concept. That is, if the agents conjecture that the economy conforms to the solution concept in the continuation and place no restrictions on the current-period matching, the set of matchings that conform to the solution concept from today onward is nonempty. 

%
%
%
%
%
%

\autoref{theorem:consistent-conjectures} suggests a way to define solution concepts for dynamic matching markets with nonempty solution correspondences by focusing attention on conjectures that satisfy consistency. We provide two illustrations of this approach in  \autoref{sec:solutions}, where we present two such conjectures and their corresponding solution concepts: \emph{continuation-value-respecting} dynamic stability and \soltwo\ \citep{hafalir2008stability}. Continuation-value-respecting dynamic stability (henceforth, \solone) is a refinement of dynamic stability \citep{doval2022dynamically}. Like the conjectures in the previous paragraph, the conjectures in \solone\ rule out continuation matchings that are not \cvr\ dynamically stable. In contrast to the conjectures in the previous paragraph, \solone\ also places restrictions on the current-period matching. In particular, each agent rules out that other contemporaneous agents leave the economy matched to matching partners that are worse than their worst conjectured matching partner. That is, each agent's conjectures respect other agents' continuation values as specified by their own conjectures. \Soltwo\ extends the solution concept in \cite{hafalir2008stability}, \emph{sophisticated expectations}, to dynamic matching markets. Sophisticated dynamic stability builds on the \emph{rational expectations solution}, a classical solution concept in matching with externalities that does not have consistent conjectures (see \autoref{example:re}). Building on the insight of \cite{hafalir2008stability}, the conjectures of sophisticated dynamic stability are iteratively built from those corresponding to rational expectations to ensure their consistency.

\paragraph{Related Literature} Our paper builds on the literature on two-sided matching with externalities \citep{li1993competitive,sasaki1996two,hafalir2008stability} to contribute to the literature on two-sided dynamic matching markets with irreversible matching \citep{doval2022dynamically,nicolo2023dynamic,nicolo2024dynamic}. 

We are closest to the contribution of \cite{hafalir2008stability}, who studies stability in static, two-sided, one-to-one, matching markets with payoff externalities. Observing the negative results in \cite{sasaki1996two} follow from conjectures being exogenously defined, \cite{hafalir2008stability} provides a sufficient condition on agents' endogenous conjectures under which stable matchings exist. As we discuss in detail in \autoref{sec:consistent-conjectures}, \autoref{theorem:consistent-conjectures} provides the analogue of \citet[Proposition 2]{hafalir2008stability} for dynamic matching markets. Whereas conceptual differences between the papers exist, the logic behind the constructions is the same. Furthermore, we provide the natural extension of the solution concept in \cite{hafalir2008stability}, sophisticated expectations, to the dynamic environments under consideration in \autoref{sec:sds}. We also introduce a new solution concept for dynamic matching markets, \solone, which has no analogue in his work.

In a static model of matching with contracts, \cite{rostek2024matching} observe that implicit assumptions on agents’ beliefs, rather than payoff externalities or complementarities, are responsible for nonexistence results. The authors show stable matchings always exist when agents’ contract choices are derived from correct beliefs about the contracts others will choose. Similarly, in \solone, agents correctly anticipate others' continuation values when deciding whether to block. However, they need not correctly anticipate others' choices in the event of a block, and existence follows from the consistency of the conjectures. 

The literature on cooperative solution concepts for dynamic matching markets highlights the role of agents' conjectures about what matching would ensue after a block  in defining solution concepts (for repeated matching markets, see, e.g., \citealp{kurino2009credibility,kadam2018multiperiod,kotowski2019perfectly,altinok2019dynamic,liu2023stability}; for irreversible matching, see, e.g., \citealp*{doval2022dynamically,nicolo2023dynamic,nicolo2024dynamic}, and for a survey treatment of this literature, see \cite{doval2023dynamic}). Because solution concepts in this literature are typically predicated on \emph{perfection} requirements---that is, the same solution concept applying at the appropriately defined continuation---agents' conjectures are defined jointly with the solution concept. We follow a complementary approach in which we take agents' conjectures as given and use them to define a solution concept. This approach allows us to identify a property of conjectures---\emph{consistency}---that guarantees the corresponding solution concept is nonempty, and underlies existence results in the literature (cf. \citealp{doval2022dynamically}). We also propose new solution concepts that refine those in the existing literature, illustrating how our result can be used as a recipe to define stability notions for dynamic matching markets.

Recently, \cite*{nicolo2024dynamic} study a third refinement of dynamic stability \citep{doval2022dynamically}. Like the conjectures in dynamic stability, the conjectures in the solution concept in \cite{nicolo2024dynamic} are defined recursively based on what the solution concept specifies in the continuation. In particular, the agents rule out all matchings that do not abide by the solution concept from tomorrow onward. Unlike dynamic stability, the solution concept in \cite{nicolo2024dynamic} imposes more structure on what matchings the agents can conjecture will ensue in the current period when considering remaining unmatched: the authors define conjectures in a given period through an iterative procedure that, at each stage, rules out current-period matchings that do not satisfy some minimal stability requirements given the conjectures at that stage.\footnote{Whereas the main focus of the paper is on the conjectures that obtain as the limit of this procedure, by stopping the procedure at a given stage the authors can study stability under different levels of sophistication \citep{hafalir2008stability}.} The conjectures they construct satisfy consistency. Following \cite{doval2022dynamically}, their existence proof is based on showing that (the analogue for their solution concept of) the candidate matchings we define in \autoref{sec:consistent-conjectures} satisfy the conditions of their solution concept. Furthermore, the authors show via an example that the rational expectations solution may fail to exist in a two-sided dynamic matching market.

Finally, our work is related to the literature on the farsighted stable set, which is used to model externalities in coalition-formation games (e.g., \citealp{harsanyi1974equilibrium,chwe1994farsighted,ray2015farsighted}), and more recently, one-to-one matching markets (e.g., \citealp{mauleon2011neumann,kimya2022farsighted}). As in this literature, agents in our model understand the terminal consequences of their moves. Whereas farsighted stability focuses on the credibility of coalitional blocks, the solution concepts we introduce in \autoref{sec:solutions} focus on the credibility of the continuation matchings used to dissuade agents from blocking.



\section{Model}\label{sec:model}
For ease of comparison, we follow the notation in \cite{doval2022dynamically}.

\paragraph{Two-sided economy} Throughout the paper, the focus is on two-sided economies, with sides \arrivala\ and \arrivalb, that last for $\terminal<\infty$ periods. Denote by \sidea\ the finite set of agents on side \arrivala, with elements $\agenta\in\sidea$. Similarly, \sideb\ denotes the finite set of agents on side \arrivalb, with elements $\agentb\in\sideb$. When we do not take a stance on an agent's side, we denote by \agent\ an element in $\sidea\cup\sideb$. 

\paragraph{Arrivals}  A length-\terminal\ economy, denoted $\economyrterminal=(\arrivala_1,\arrivalb_1,\dots,\arrivala_\terminal,\arrivalb_\terminal)$, consists of two sequences, $\{\arrivala_1,\dots,\arrivala_\terminal\}$ and $\{\arrivalb_1,\dots,\arrivalb_\terminal\}$, of disjoint subsets of \sidea\ and \sideb, respectively.\footnote{Without loss, we can assume the sets $(\arrivala_1,\dots,\arrivala_\terminal)$ are disjoint because we do not rule out indifferences. Thus, \agenta\ denotes a unique individual in the economy, but \agenta\ may be payoff equivalent to other arriving agents on side \arrivala. The same applies for the sets $(\arrivalb_1,\dots,\arrivalb_\terminal)$ being disjoint.}  For $t\in\{1,\dots,\terminal\}$,  \arrivalat\ are the side-\arrivala\ agents that arrive in period $t$, and \arrivalbt\ are the side-\arrivalb\ agents that arrive in period $t$. For any $t\leq\terminal$, let $\arrivalauptot=\cup_{\dateindex=1}^t\arrivala_\dateindex$ denote the implied arrivals on side \arrivala\ through period $t$; similarly, let $\arrivalbuptot=\cup_{\dateindex=1}^t\arrivalb_\dateindex$ denote the implied arrivals on side \arrivalb\ through period $t$.

\paragraph{Dynamic Matching} With this notation at hand, we define matchings in a given period (\autoref{definition:static-matching}) and matchings for the dynamic economy (\autoref{definition:matching}):
\begin{definition}[Period-$t$ matching]\label{definition:static-matching}
A \emph{period-$t$ matching} is a mapping
$\matchingt:\arrivalauptot\cup\arrivalbuptot\rightarrow\arrivalauptot\cup\arrivalbuptot$ such that the following hold:
\begin{enumerate}
    \item For all $\agenta\in\arrivalauptot$, $\matchingt(\agenta)\in\{\agenta\}\cup\arrivalbuptot$,
    \item For all $\agentb\in\arrivalbuptot$, $\matchingt(\agentb)\in\arrivalauptot\cup\{\agentb\}$,
    \item For all $\agent\in\arrivalauptot\cup\arrivalbuptot$, $\matchingt(\matchingt(\agent))=\agent$.
\end{enumerate}
\end{definition}
\begin{definition}[Irreversible dynamic matching]\label{definition:matching}
A \emph{matching} \matching\ is a tuple $(\matching_1,\dots,\matching_\terminal)$ such that
\begin{enumerate}
    \item\label{itm:t-matching} For all $t\in\{1,\dots,\terminal\}$, $\matchingt$ is a period-$t$ matching,
    \item\label{itm:irreversible} For all $t\in\{1,\dots,\terminal\}$, for all $\agenta\in\arrivalauptot$, if $\matchingt(\agenta)\neq \agenta$, then $\matching_\dateindex(\agenta)=\matchingt(\agenta)$ for all $\dateindex\geq t$.
\end{enumerate}
Let \Matchingsfinal\ denote the set of matchings. 
\end{definition}
Part \ref{itm:irreversible} of the definition guarantees that once a pair \pair\ is matched, they are matched in all subsequent periods. Given a matching $\matching\in\Matchingsfinal$, let \matchinguptot\ denote the tuple $(\matching_1,\dots,\matching_{t-1})$, with $\matching^0=\{\emptyset\}$ and let $\Matchingsfinal(\matchinguptot)$ denote the set of matchings that coincide with \matching\ through period $t-1$. For an agent $\agent\in\arrivalauptot\cup\arrivalbuptot$, let $\Matchingsfinal(\matchinguptot,\agent)$
denote the subset of $\Matchingsfinal(\matchinguptot)$ such that \agent\ is unmatched through period $t$.

 \paragraph{Available agents and induced economies} Fix a matching $\matching\in\Matchingsfinal$, and suppose agents have matched according to \matching\ through period $t-1$. Then, the set of agents who can match in period $t$ is determined by the unmatched agents in period $t-1$ and the new arrivals in period $t$, $(\arrivala_t,\arrivalb_t)$. Formally,
\begin{align*}
    \cala_t(\matchinguptot)&=\{\agenta\in\arrivalaupto_{t-1}:\matching_{t-1}(\agenta)=\agenta\}\cup\arrivala_t,\\
    \calb_t(\matchinguptot)&=\{\agentb\in\arrivalbupto_{t-1}:\matching_{t-1}(\agentb)=\agentb\}\cup\arrivalb_t.
\end{align*}
For each $t\geq1$, a matching \matching\ induces a continuation economy of length $\terminal-t$, $\economyr_{t+1}^\terminal(\matching^t)=(\cala_{t+1}(\matching^t),\calb_{t+1}(\matching^t),\dots,\arrivala_\terminal,\arrivalb_\terminal)$. Given a matching $\matchingb\in\Matchingsfinal(\matching^t)$, we denote  the restriction of \matchingb\ to the continuation economy $\economyr_{t+1}^\terminal(\matching^t)$  by $(\matchingb_s)_{s= t+1}^\terminal|_{\economyr_{t+1}^\terminal(\matching^t)}$.\label{page-r3-notation}

\paragraph{Preferences} We close the model by defining the agents' preferences.
Throughout, we assume agents discount the future. Formally, each $\agenta\in\sidea$ defines a discount factor $\delta_\agenta\in[0,1]$ and a Bernoulli utility, $u(\agenta,\cdot):\sideb\cup\{\agenta\}\rightarrow\mathbb{R}$. Similarly, each $\agentb\in\sideb$ defines a discount factor $\delta_\agentb\in[0,1]$ and a Bernoulli utility, $v(\cdot,\agentb):\sidea\cup\{b\}\rightarrow\mathbb{R}$.\footnote{We follow the convention that $0^0=1$.\label{page-r2-discount}\label{fnt:discount}} We normalize to $0$ the payoffs of remaining single for one period. That is, for all $\agenta\in\sidea$ and all $\agentb\in\sideb$, $u(\agenta,\agenta)=v(\agentb,\agentb)=0$. 
 
 Given a matching $\matching\in\Matchingsfinal$, let $U_t(\agenta,\matching)$ and $V_t(\agentb,\matching)$ denote respectively \agenta's and \agentb's utilities of matching \matching\ from period $t$ onwards. Formally, fix a matching $\matching\in\Matchingsfinal$ and a period $t$. For every $\agent\in\cala_t(\matchinguptot)\cup\calb_t(\matchinguptot)$, let $t_\matching(\agent)$ denote the first date at which \agent\ is matched under \matching. That is, $t_\matching(\agent)$ is the smallest index \dateindex\ such that $t\leq\dateindex$ and $\matching_\dateindex(\agent)\neq \agent$; otherwise, let $t_\matching(\agent)=\terminal$. Then, for $\agenta\in\cala_t(\matchinguptot)$, let
\begin{align*}
U_t(\agenta,\matching)=\delta_a^{t_\matching(a)-t}u(a,\matching_\terminal(a)),
\end{align*}
denote \agenta's payoff from matching \matching\ at date $t$. Similarly, for $\agentb\in\calb_t(\matchinguptot)$, let
\begin{align*}
V_t(\agentb,\matching)=\delta_b^{t_{\matching}(b)-t}v(\matching_\terminal(b),b),
\end{align*}
denote \agentb's payoff from matching \matching\ at date $t$.

We record for future reference two properties a matching \matching\ may satisfy:
\begin{definition}[Individual rationality]\label{definition:ir}
A matching $\matching\in\Matchingsfinal$ is \emph{individually rational} if for all $\agenta\in\sidea$, $u(\agenta,\matching_\terminal(\agenta))\geq0$, and for all $\agentb\in\sideb$, $v(\matching_\terminal(\agentb),\agentb)\geq0$.\end{definition}

\begin{definition}[Static notion of stability]\label{definition:galeshapley} Suppose $\terminal=1$. A matching $\matching$ for economy $\economyr^1=(\arrivala_1,\arrivalb_1)$ is \emph{stable} if the following hold:
\begin{enumerate}[leftmargin=*]
\item \matching\ is individually rational, and 
\item no pair $\pair\in\arrivala_1\times\arrivalb_1$ exists such that $\genericu>U_1(\agenta,\matching)$ and $\genericv>V_1(\agentb,\matching)$.
\end{enumerate}
We denote by $\stable(\economyr^1)$ the set of stable matchings for $\economyr^1$.
\end{definition}



\begin{remark}[Notation]\label{remark:notation} We note several pieces of notation, such as \Matchingsfinal, $\Matchingsfinal(\matchinguptot)$, $\cala_t(\cdot)$, $\calb_t(\cdot)$, denote objects whose definitions depend on the economy \economyrterminal\ under consideration. To be precise, we should note them: $\Matchingsfinal(\economyrterminal),\Matchingsfinal(\matchinguptot;\economyrterminal),\calat(\cdot;\economyrterminal),\calbt(\cdot;\economyrterminal)$. In what follows, we continue to omit this dependence whenever doing so does not lead to confusion.
\end{remark}
\begin{remark}[Simplifying assumptions]\label{remark:ass} We note two assumptions that simplify notation, but are otherwise not needed for the results. First, we assume agents can perfectly foresee who arrives in each period. Our results extend to the case in which arrivals are stochastic so that an economy is defined as a distribution over sequences \economyrterminal\ as in \cite{doval2020dynamically}. Second, we assume time-discounted preferences, but all that matters for our results is that agents are dynamically consistent.
\end{remark}

\section{A solution concept given conjectures}\label{sec:conjectures}
In this section, we introduce the two key objects of our paper: an agent's conjectures and a solution concept induced by these conjectures. 

%
%

Fix an economy \economyrterminal, a matching \matching, a period $t$, and an agent \agent\ who is available to match in period $t$. Agent \agent's conjectures describe the matchings \agent\ considers possible should they decide to remain unmatched in period $t$:
\label{page-conjectures}\begin{definition}[Conjectures]\label{definition:conjectures}
Fix a period $t$, a matching through $t-1$, \matchinguptot, and an agent $\agent\in\cala_t(\matchinguptot)\cup\calb_t(\matchinguptot)$. A conjecture for agent \agent\ is a nonempty subset $\conjecturet(\matchinguptot,\agent)$ of $\Matchingsfinal(\matchinguptot,\agent)$.
\end{definition}
Because the conjectures represent what \agent\ expects to obtain should they remain unmatched in period $t$, each conjectured matching \matchingb\ satisfies $\matchingbt(\agent)=\agent$. Each conjectured matching $\matchingb\in\conjecturet(\matchinguptot,\agent)$ describes both the matching that ensues in period $t$, \matchingbt, and the continuation matching $(\matchingb_{t+1},\dots,\matchingb_\terminal)$. Whereas agent \agent's matching outcome is determined by the continuation matching  $(\matchingb_{t+1},\dots,\matchingb_\terminal)$, specifying who matches in period $t$ is necessary to define what matchings are feasible from period $t+1$ onward. For this reason, a conjecture describes the matching starting from period $t$ onward. That defining the set of matchings \agent\ may face from period $t+1$ onward necessitates specifying the outcomes of the other period-$t$ agents implies an externality exists in this economy, even if no direct payoff externalities exist.
%


Given a family of conjectures, $\conjectures\equiv\{\conjecturet(\matchinguptot,\agent):1\leq t\leq\terminal,\matchinguptot,\agent\in\sidea\cup\sideb\}$, one can define a solution concept for \economyrterminal, as follows:

\begin{definition}[Solutions induced by \conjectures]\label{definition:sol-given-c}
A matching $\matching\in\Matchingsfinal$  is a \conjectures-solution for economy \economyrterminal\ if for all $t\geq1$ the following hold:
\begin{enumerate}
\item\label{itm:ira} For all $\agenta\in\calat(\matchinguptot)$, a matching $\matchingb\in\conjecturet(\matchinguptot,\agenta)$ exists such that 
$U_t(\agenta,\matching)\geq U_t(\agenta,\matchingb)$,
\item\label{itm:irb} For all $\agentb\in\calbt(\matchinguptot)$, a matching $\matchingb\in\conjecturet(\matchinguptot,\agentb)$ exists such that 
$V_t(\agentb,\matching)\geq V_t(\agentb,\matchingb)$,
\item\label{itm:ps} No pair $\pair\in\calat(\matchinguptot)\times\calbt(\matchinguptot)$ exists such that 
$\genericu>U_t(\agenta,\matching)\text{ and }\genericv>V_t(\agentb,\matching)$.
\end{enumerate}
Let $\sol_\terminal^\Phi(\economyrterminal)$ denote the set of \conjectures-solutions for \economyrterminal.
\end{definition}
In words, a matching \matching\ is a \conjectures-solution for \economyrterminal\ if in each period the following holds. First, each agent \agent\ who is available to match in period $t$ prefers \matching\ to their worst conjectured matching (parts \ref{itm:ira} and \ref{itm:irb} in \autoref{definition:sol-given-c}). Note this condition implies that if $\matchingt(\agent)\neq\agent$, \agent\ prefers their matching partner to waiting to be matched, and if $\matchingt(\agent)=\agent$, \agent\ has no objections to the continuation matching, $(\matching_s)_{s=t+1}^\terminal$.\footnote{In a static economy, only the first kind of block may exist. The second block only shows up in a dynamic economy (see \citealp{doval2022dynamically}, for a discussion). For instance, if the continuation matching is not self-enforcing, an agent may object to it. As we discuss later, consistency of the conjectures is related to the second class of blocks.} In both cases, we say \agent\ prefers to be available to match in period $t$. Second, no pair of agents who can match in period $t$ prefer to match with each other to matching according to \matching\ (part \ref{itm:ps} in \autoref{definition:sol-given-c}). Note \pair\ being able to match in period $t$ does not imply \agenta\ and \agentb\ arrive in the same period, but that under matching \matching, they meet in period $t$ and are able to block \matching. 

\autoref{definition:sol-given-c} is in the spirit of the static notion of stability: \conjectures-solutions are defined by the absence of pairwise blocks and the requirement that each agent is matched to a partner that is preferred to remaining unmatched.\footnote{\autoref{definition:sol-given-c} is also similar to that of dynamic stability in \cite{doval2022dynamically}, except the conjectures in dynamic stability are recursively defined using the solution concept. In other words, \autoref{definition:sol-given-c} nests that of dynamic stability by allowing for arbitrary conjectures.} In contrast to the static notion of stability, the value of remaining unmatched is determined by the conjectures.  Indeed, when $\terminal=1$, the correspondence $\sol_1^\conjectures$ reduces to the set of stable matchings (\autoref{definition:galeshapley}). To illustrate, consider a length-1 economy, $\economyr^1=(\arrivala_1,\arrivalb_1)$, a matching \matching\ for $\economyr^1$, and an agent $\agent\in\arrivala_1\cup\arrivalb_1$. Because all matchings $\matchingb\in\conjecture_1(\emptyset,\agent)$ are such that $\matchingb_1(\agent)=\agent$, parts \ref{itm:ira} and \ref{itm:irb} of \autoref{definition:sol-given-c} simply state that \agent\ prefers \matching\ to remaining single. That is, \matching\ is individually rational. Furthermore, part \ref{itm:ps} implies \matching\ has no pairwise blocks. Thus, when $\terminal=1$, \autoref{definition:sol-given-c} reduces to \autoref{definition:galeshapley} for any family of conjectures. We record this observation for future reference:
\begin{observation}\label{observation:stability}
For all economies of length $1$, $\economyr^1$, and for all conjectures, \conjectures, the set of \conjectures-solutions for $\economyr^1$ corresponds to the set of stable matchings for $\economyr^1$. That is, for all $\economyr^1$ and all $\conjectures$, $\sol_1^\conjectures(\economyr^1)=\stable(\economyr^1)$.
\end{observation}
Among other things, \autoref{observation:stability} implies \conjectures-solutions are always nonempty for economies of length $1$. We use this property repeatedly in our proofs.

Whereas \autoref{definition:sol-given-c} is fairly general, note the definition rules out solution concepts in which the conjectures depend on the matching under consideration beyond the induced economy in period $t$. For instance, when considering whether to block matching \matching\ in period $t$, if agent \agent\ presumes all other contemporaneous agents\textemdash except perhaps for \agent's matching partner\textemdash remain matched according to \matchingt, \agent's conjectures depend on the outcome specified by \matching\ in period $t$. (These conjectures are referred to as \emph{passive} conjectures; see \cite{chowdhury2004marriage} and \cite{doval2015theory}.) Instead, \autoref{definition:sol-given-c} presumes we can define the conjectures without reference to any specific matching. The matching under consideration only matters inasmuch as it determines who matches in each period, and hence whose conjectures one should check in each period.

%
%

We close this section by providing a recursive definition of the set of \conjectures-solutions for a length-\terminal\ economy, which simplifies the formal statements in the following sections: 
\begin{definition}[Recursive]\label{definition:recursive} Given the correspondences $(\sol_t^\conjectures)_{t=1}^{\terminal-1}$, matching $\matching\in\Matchingsfinal$ is a \conjectures-solution for economy \economyrterminal\ if the following hold:
\begin{enumerate}
\item\label{itm:ira-r} For all $\agenta\in\arrivala_1$, a matching $\matchingb\in\conjecturebeg(\emptyset,\agenta)$ exists such that 
$U_1(\agenta,\matching)\geq U_1(\agenta,\matchingb)$,
\item\label{itm:irb-r} For all $\agentb\in\arrivalb_1$, a matching $\matchingb\in\conjecturebeg(\emptyset,\agentb)$ exists such that 
$V_1(\agentb,\matching)\geq V_1(\agentb,\matchingb)$,
\item\label{itm:ps-r} No pair $\pair\in\arrivala_1\cup\arrivalb_1$ exists such that 
$
\genericu>U_1(\agenta,\matching)\text{ and }\genericv>V_1(\agentb,\matching)$, and
\item\label{itm:cont} $(\matching_t)_{t=2}^\terminal|_{\economyrterminal_2(\matching_1)}\in\sol_{\terminal-1}^\conjectures(\economyr_2^\terminal(\matching_1))$.
\end{enumerate}
\end{definition} 
\autoref{definition:recursive} simply states matching \matching\ is a \conjectures-solution for $\economyr^\terminal$ if the following two conditions hold. First, from period $2$ onwards, \matching\ specifies a \conjectures-solution for the induced economy of length $\terminal-1$, $\economyr_2^\terminal(\matching_1)$. Second, \matching\ has no  blocks in period $1$ in the sense of \autoref{definition:sol-given-c}. Because the correspondence $\sol_{\terminal-1}^\conjectures$ depends only on the conjectures from period $2$ onward, this recursive definition allows us in what follows to focus on the properties of $\conjecturebeg(\emptyset,\cdot)$, taking as given that the set $\sol_{\terminal-1}^\conjectures$ is nonempty.

\section{Consistent conjectures}\label{sec:consistent-conjectures}
In this section, we present a sufficient condition on the family of conjectures, \conjectures, such that the set of \conjectures-solutions is nonempty (\autoref{theorem:consistent-conjectures}). We dub the conjectures satisfying the sufficient condition as \emph{consistent} conjectures. Consistent conjectures guarantee candidate matchings, defined below, are a solution. In what follows, we first introduce the candidate matchings, and then we state \autoref{theorem:consistent-conjectures}. After discussing the intuition behind the result, we connect it to similar observations in \cite{hafalir2008stability} and \cite{doval2022dynamically}. We conclude this section by showing  the extension to dynamic matching markets of a standard solution concept in matching with externalities, rational expectations, does not entail consistent conjectures.

\paragraph{Candidate matchings} Given an economy \economyrterminal\ and the family of conjectures \conjectures, we introduce a procedure to construct a set of natural candidates to be \conjectures-solutions for \economyrterminal.  This procedure uses the conjectures to transform the dynamic economy into a sequence of static economies \emph{without} externalities. The next definition is key:

\begin{definition}[One-period economy induced by $\conjecturebeg$]\label{definition:one-shot-economy}
Given an economy \economyrterminal\ and a matching $\matching\in\Matchingsfinal$, the one-period economy induced by the collection $\{\conjecturebeg(\emptyset,\agent):\agent\in\arrivala_1\cup\arrivalb_1\}$ is the length-1 economy defined as follows:
\begin{enumerate}
    \item Agents are $\arrivala_1\cup\arrivalb_1$,
    \item Utilities  $\{\payoffacb(\agenta,\cdot):\arrivalb_1\cup\{\agenta\}\rightarrow\mathbb{R}|\agenta\in\arrivala_1\},\{\payoffbcb(\cdot,\agentb):\arrivala_1\cup\{\agentb\}\rightarrow\mathbb{R}\vert\agentb\in\arrivalb_1\}$ are given by:
    \begin{enumerate}
\item    For each agent $\agenta\in\arrivala_1$, $\payoffacb(\agenta,\agenta)=\min\{U_1(\agenta,\matchingb):\matchingb\in \conjecturebeg(\emptyset,\agenta)\}$. Moreover, $\payoffacb(\agenta,\agentb)=\genericu$ for $\agentb\in\arrivalb_1$.
  
       \item  For each agent $\agentb\in\arrivalb_1$, $\payoffbcb(\agentb,\agentb)=\min\{V_1(\agentb,\matchingb):\matchingb\in \conjecturebeg(\emptyset,\agentb)\}$. Moreover, $\payoffbcb\pair=\genericv$ for $\agenta\in\arrivala_1$.
       \end{enumerate}
\end{enumerate}
We denote by $\economyr_{1,\conjecturebeg}$ the one-period economy induced by \conjecturebeg.
\end{definition}
In the one-period economy induced by the conjectures $\conjecturebeg$, an agent \agenta\ ranks matching partners on side \arrivalb\ according to $u(\agenta,\cdot)$, except that agents on side \arrivalb\ who are worse than the worst conjectured continuation matching are now deemed unacceptable. In other words, in the one-period economy induced by $\conjecturebeg$, each agent in $\arrivala_1\cup\arrivalb_1$ has the worst conjectured matching as an outside option.

The construction in \autoref{definition:one-shot-economy} can be understood from the point of view of both matching with externalities and dynamic programming. On the one hand, \cite{sasaki1996two} show how to use a set of conjectures to construct an economy without externalities (we discuss this construction formally after \autoref{theorem:consistent-conjectures}). \autoref{definition:one-shot-economy} builds on their insight: we use the conjectures \conjecturebeg\ to define each period-1 agent's value from remaining unmatched \emph{independently} of other contemporaneous agents' matching outcomes. Thus, no externalities exist in the economy $\economyr_{1,\conjecturebeg}$. On the other hand, \autoref{definition:one-shot-economy} is also reminiscent of the logic behind dynamic programming: We can reduce a dynamic economy to a static one by appropriately specifying each agent's continuation values, which in this case are given by their value of remaining unmatched as defined by \conjecturebeg.

Suppose we know $\sol_{\terminal-1}^\conjectures$ is a nonempty-valued correspondence and consider the following set of \emph{candidate} matchings:
\begin{align*}
\candidate(\conjecturebeg,\sol_{\terminal-1}^\conjectures)=\left\{\matchingstar\in\Matchingsfinal:\matchingstar_1\in\stable\left(\economyr_{1,\conjecturebeg}\right)\text{ and } (\matchingstar_t)_{t=2}^\terminal|_{\economyr_2^\terminal(\matchingstar_1)}\in\sol_{\terminal-1}^\conjectures(\economyr_2^\terminal(\matchingstar_1))\right\}.
\end{align*}

In words, a candidate matching, $\matchingstar\in\candidate$, satisfies the following. In period $1$, $\matching_1^\star$ is a stable matching for $\economyr_{1,\conjecturebeg}$, that is, $\matchingstar_1\in\stable\left(\economyr_{1,\conjecturebeg}\right)$. Because the one-period economy induced by $\conjecturebeg$ is devoid of any externalities, its set of stable matchings is nonempty, and hence, at least one such matching exists \citep{gale1962college}. Given the period-1 matching, $\matchingstar_1$, the continuation matching, $(\matchingstar_t)_{t=2}^\terminal$ is selected from $\sol_{\terminal-1}^\conjectures(\economyr_2^\terminal(\matchingstar_1))$, which is nonempty by assumption. The set of candidate matchings may not be a singleton either because $\stable\left(\economyr_{1,\conjecturebeg}\right)$ is not a singleton, or because, for a given stable matching in $\stable\left(\economyr_{1,\conjecturebeg}\right)$, multiple continuation matchings may be solutions in the induced economy.\footnote{Whereas many stable matchings for the one-period economy induced by the period-1 conjectures may exist, the Lone Wolf Theorem \citep{mcvitie1970stable} implies they all have the same set of unmatched agents, and hence, they all induce the same continuation economy.\label{fnt:lone-wolf}}

Any matching in $\candidate(\conjecturebeg,\sol_{\terminal-1}^\conjectures)$ is a natural candidate to be a \conjectures-solution because it satisfies most of the conditions in \autoref{definition:recursive}. To illustrate, fix any such matching, \matchingstar, and note the following. First, \matchingstar\ satisfies part \ref{itm:cont} of \autoref{definition:recursive} by construction. Second, \matchingstar\ satisfies parts \ref{itm:ira-r}, \ref{itm:irb-r}, and \ref{itm:ps-r} of \autoref{definition:recursive} for all period-1 agents $\agent\in\arrivala_1\cup\arrivalb_1$ such that $\matchingstar_1(\agent)\neq\agent$. By definition of $\economyr_{1,\conjecturebeg}$, every agent who matches in \periodone\ is matched to someone better than their worst conjectured matching. Furthermore, because $\matchingstar_1$ is chosen from $\stable\left(\economyr_{1,\conjecturebeg}\right)$, no pair of matched agents under $\matchingstar_1$ can form a blocking pair.

\paragraph{Toward consistent conjectures} As we explain next, the only reason any such matching \matchingstar\ may fail to be a \conjectures-solution is that a ``gap'' exists between the conjectures of an agent who is unmatched in \periodone\ under \matchingstar\ and  the matching \matchingstar. To illustrate, note the above discussion implies that if \matchingstar\ is not a \conjectures-solution, \matchingstar\ has a block involving a period-$1$ agent, \agent, who is unmatched in \periodone, that is, $\matchingstar_1(\agent)=\agent$. We claim next that if \agent\ blocks \matchingstar, then \matchingstar\ is not a conjecture for \agent. To fix ideas, let $\agent=\agenta\in\arrivala_1$ and suppose \agenta\ blocks $\matchingstar_1$ with an agent $\agentb\in\arrivalb_1$. Because $\matchingstar_1\in\stable\left(\economyr_{1,\conjecturebeg}\right)$, it must be that at least one of \agenta\ or \agentb\ prefers their worst conjectured matching to matching with each other; assume for simplicity this holds for \agenta. Because \agenta\ wants to block \matchingstar\ with \agentb, it must be that \agenta's outcome under \matchingstar\ is \emph{worse} than the worst matching in $\conjecturebeg(\emptyset,\agenta)$. In other words, \matchingstar\ is not an element of agent \agenta's conjectures.   Because \agenta\ cares only about the continuation matching $(\matchingstar_t)_{t=2}^\terminal$, that \matchingstar\ is not part of \agenta's conjectures is akin to agent \agenta's conjectures not being \emph{consistent} with the solution concept in the continuation. If, instead, \agenta's conjectures were consistent with the solution concept in the continuation, they would not be willing to block \matchingstar\ with \agentb.

The above discussion motivates the following definition:

\label{page-cc}\begin{definition}[Consistent conjectures]\label{definition:consistent-conjectures}
Suppose $\sol_{\terminal-1}^\conjectures$ is a nonempty-valued correspondence and fix an economy \economyrterminal. Conjectures $\{\conjecturebeg(\emptyset,\agent):\agent\in\arrivala_1\cup\arrivalb_1\}$ are consistent if for all candidate matchings $\matchingstar\in\candidate(\conjecturebeg,\sol_{\terminal-1}^\conjectures)$, the following holds: 
\begin{align}\label{eq:cc}\tag{CC}
\text{for all $\agent\in\arrivala_1\cup\arrivalb_1$ such that $\matchingstar_1(\agent)=\agent$, we have that }\matchingstar\in\conjecturebeg(\emptyset,\agent).\end{align} 
\end{definition}
In words, conjectures are consistent if each candidate matching, \matchingstar, is common to the conjectures of all agents who are unmatched under \matchingstar\ in period $1$. For a given candidate matching \matchingstar, property \ref{eq:cc} places a constraint on the conjectures of those agents who are unmatched in period $1$ under \matchingstar: all of them should agree  $(\matchingstar_t)_{t=2}^\terminal$ is a valid continuation matching. Instead, property \ref{eq:cc} places no restriction on those agents who are matched in period $1$ under \matchingstar: indeed, these agents may have different conjectures about the matching that would ensue conditional on remaining unmatched.
%
%
%

We have the following result:
\begin{theorem}[Solutions induced by consistent conjectures are nonempty]\label{theorem:consistent-conjectures}
Suppose $\sol_{\terminal-1}^\conjectures$ is a nonempty-valued correspondence. Fix \economyrterminal\ 
and suppose the conjectures $\{\conjecturebeg(\emptyset,\agent):\agent\in\arrivala_1\cup\arrivalb_1\}$ are consistent. Then, $\sol_\terminal^\conjectures(\economyrterminal)$ is nonempty. In particular,  $\candidate(\conjecturebeg,\sol_{\terminal-1}^\conjectures)\subset\sol_\terminal^\conjectures(\economyrterminal)$.
\end{theorem}
In words, consistency of the conjectures guarantees each of the candidate matchings are solutions for the economy. The proof of \autoref{theorem:consistent-conjectures}, which can be found in \autoref{appendix:definitions}, shows a stronger result: if property \ref{eq:cc} holds for \emph{some} candidate matching, \matchingstar, then  \matchingstar\ is a solution for the economy. That is, to show the set of \conjectures-solutions for \economyrterminal\ is nonempty, assuming property \ref{eq:cc} holds for \emph{some} candidate matching is enough. 
Because the nonempty-valued solution concepts we introduce in this and the following section satisfy the stronger property in \autoref{definition:consistent-conjectures}, we prefer to make that our definition of consistency.


%

As stated, the condition in \autoref{theorem:consistent-conjectures} seems difficult to check without actually computing the set of candidate matchings. Yet,
\autoref{theorem:consistent-conjectures} suggests conjectures that always lead to nonempty solutions \emph{without} actually computing \matchingstar. Indeed, it follows from \autoref{theorem:consistent-conjectures} that if each period-1 agent conjectures that from \periodtwo\ onward, the continuation matching is an element of $\sol_{\terminal-1}^\conjectures(\cdot)$, $\sol_\terminal^\conjectures$ is nonempty valued. Formally, define for each $\agent\in\arrivala_1\cup\arrivalb_1$, 
\begin{align}\label{eq:eqbm-cont}
\conjecturebeg^*(\emptyset,\agent)=\left\{\matchingb\in\Matchingsfinal(\emptyset,\agent):(\matchingb_t)_{t=2}^\terminal|_{\economyr_2^\terminal(\matchingb_1)}\in\sol_{\terminal-1}^\conjectures(\economyr_2^\terminal(\matchingb_1))\right\}.
\end{align}
These conjectures reflect agent \agent\ believes the economy abides by the set of \conjectures-solutions from period $2$ onward, akin to a perfection requirement. For instance, if $\terminal=2$, \autoref{eq:eqbm-cont} states agent \agent\ anticipates the continuation matching, $\matchingb_2$, is a stable matching in the induced economy, $\economyr_2^2(\matchingb_1)$. That these conjectures induce a nonempty solution concept has two interesting implications. First, considering larger conjectures (in the set-inclusion sense) than those in \autoref{eq:eqbm-cont} is not needed to guarantee existence. The latter is a desirable property because a larger set of conjectures than those in \autoref{eq:eqbm-cont} would include matchings whose continuations do not satisfy the solution concept and are, in a sense, non-credible. Second, it implies the perfection requirement alone is compatible with the existence of solutions for dynamic matching markets. Instead, difficulties may arise if the conjectures also place structure on the period-$t$ matching, as we do in \autoref{sec:solutions}.

Even if the statement of \autoref{theorem:consistent-conjectures} relies on the recursive definition of the set of \conjectures-solutions for a length-\terminal\ economy taking the nonemptiness of $\sol_{\terminal-1}^\conjectures$ as given, the procedure to construct \matchingstar\ and the statement of \autoref{theorem:consistent-conjectures} can be expressed solely in terms of the model primitives. We do this in \autoref{appendix:definitions} for completeness. When expressed in terms of the model primitives, consistency requires property \ref{eq:cc} holds in each period along the path of \matchingstar. That is, consistency is defined as the requirement that all candidate matchings, \matchingstar, satisfy that for all $t\geq1$, all $\agent\in\calat(\matchingstaruptot)\cup\calbt(\matchingstaruptot)$ such that $\matchingstar_t(\agent)=\agent$, then $\matchingstar\in\conjecture_t(\matchingstaruptot,\agent)$ (see \autoref{definition:cc-app}). In the same appendix, we formally define the conjectures and the solution concept suggested by  \autoref{eq:eqbm-cont}. That is, we iteratively apply \autoref{eq:eqbm-cont}, starting from length-1 economies, to define a solution concept with conjectures consisting of all matchings whose continuations are solutions in the induced continuation economy. By \autoref{theorem:consistent-conjectures}, this solution concept is always nonempty (see \autoref{corollary:agree} in \autoref{appendix:definitions}).\footnote{The solution concept defined in \autoref{appendix:definitions} is the analogue for irreversible matching markets of the one introduced in \cite{kotowski2019perfectly} for repeated matching markets.\label{fnt:kotowski}}

%

We close this section by (i) discussing the connection with \cite{hafalir2008stability} and \cite{doval2022dynamically},  and why consistent conjectures are conceptually appealing, and (ii) illustrating through an example that \emph{rational expectations}, a traditional solution concept in matching with externalities, does  not have consistent conjectures.

\textbf{Connection with \protect{\cite{hafalir2008stability}} and \protect{\cite{doval2022dynamically}}} \cite{hafalir2008stability} studies stability in two-sided, one-to-one, matching markets with payoff externalities.
The starting point of his analysis is that the negative results in \cite{sasaki1996two} follow from conjectures being exogenously defined without regard to the economy at hand. The first main result in \cite{hafalir2008stability} provides a sufficient condition on the conjectures for a stable matching to exist in his environment. To explain the condition, we review the construction in \cite{hafalir2008stability}. Using the conjectures, \cite{hafalir2008stability} constructs a two-sided economy without externalities as in \cite{sasaki1996two}. In present notation, consider a one-period economy with agents in $\arrivala_1\cup\arrivalb_1$. For each pair of agents, $\pair\in\arrivala_1\times\arrivalb_1$, a conjecture $\conjecture\pair$ is a subset of the set of matchings such that \pair\ is matched.\footnote{\cite{hafalir2008stability} does not consider blocks by individual agents.} Given the conjectures, say \agenta\ prefers \agentb\ to \agentbb\ if \agenta\ prefers the worst matching in \conjecture\pair\ to the worst matching in $\conjecture(\agenta,\agentbb)$. The preferences for agents on side \arrivalb\ are similarly defined. Under these induced preferences, a stable matching is guaranteed to exist. Fix any such matching, \matchingb. \citet[Proposition 2]{hafalir2008stability} shows that if for all pairs \pair\ such that $\matchingb(\agenta)=\agentb$, we have that $\matchingb\in\conjecture\pair$, then \matchingb\ is stable in the economy with externalities. 

\autoref{theorem:consistent-conjectures} extends the logic of the result in \cite{hafalir2008stability} to dynamic matching markets. In contrast to \cite{hafalir2008stability}, no payoff externalities exist in our model; that is, an agent \agent's payoff from a matching \matching\ depends only on the agent's matching partner, $\matching_\terminal(\agent)$. Instead, other agents' outcomes are relevant to define \agent's payoff from remaining unmatched in a given period. For this reason,  \autoref{theorem:consistent-conjectures} only requires that consistency holds for those agents who are unmatched under \matchingstar, whereas the condition in \cite{hafalir2008stability} must hold for all matched pairs. 


\label{page-r1-doval}\autoref{theorem:consistent-conjectures} identifies a key property that underlies the existence result in \cite{doval2022dynamically}. In fact, Remark 2 in that paper points out that the conjectures corresponding to dynamic stability satisfy condition \ref{eq:cc} for any dynamically stable matching---not just the corresponding candidate matchings \matchingstar---and discusses the key role this property plays in the existence result. In \cite{doval2022dynamically}, the conjectures are defined recursively based on what the solution concept specifies in the continuation, similar to the conjectures in \autoref{eq:eqbm-cont}. Because our framework isolates the conjectures from the solution concept, it allows us to identify the role of consistency without reference as to how the conjectures relate to the solution concept. In doing so, the result offers an alternative way to construct solution concepts for dynamic matching markets by focusing on guaranteeing the conjectures are consistent. \autoref{sec:solutions} illustrates this approach.

\paragraph{Why consistent conjectures?} Because \autoref{theorem:consistent-conjectures} offers a recipe of sorts to construct solution concepts in dynamic matching markets  with nonempty predictions, considering the conceptual merits of consistent conjectures is natural. We consider two in what follows: the first builds on a similar argument in \cite{hafalir2008stability}; the second expands on the notion that consistency is akin to the agents understanding and agreeing that the solution concept represents the set of self-enforcing matchings in the continuation economy.
%

First,  considering the implications of not having consistent conjectures in the context of the candidate matchings is instructive. To fix ideas, consider one such \matchingstar\ and suppose an unmatched agent on side \arrivala, \agenta, wishes to block \matchingstar\ with an agent on side \arrivalb, \agentb. Suppose also that \agenta\ prefers the worst matching in $\conjecturebeg(\emptyset,\agenta)$ to matching with \agentb, and prefers matching with agent \agentb\ to \matchingstar, so that $\matchingstar\notin\conjecturebeg(\emptyset,\agenta)$. Agent \agenta's block is in a sense not credible: whereas \agenta\ prefers matching with \agentb\ to \matchingstar, as soon as \agenta\ matches with \agentb, \agenta\ would prefer to break up with \agentb. After all, \agenta\ prefers the worst matching in 
$\conjecturebeg(\emptyset,\agenta)$ to matching with \agentb. Thus, whenever the failure of consistency rules out a candidate matching \matchingstar\ as a solution, this may be because the solution concept allows for these non-credible blocks.

To make the second point, considering a more general definition of consistency that dispenses with the candidate matchings is useful. Given a family of conjectures, \conjectures, and the induced solution concept $\{\sol_t^\conjectures:t\in\{1,\dots,\terminal\}\}$, we say the solution concept satisfies \emph{generalized consistency} if the following holds. For all economies \economyrterminal\ and \conjectures-solutions \matching\ for \economyrterminal, for all periods $t$, and $\agent\in\calat(\matchinguptot)\cup\calbt(\matchinguptot)$, if $\matchingt(\agent)=\agent$, then $\matching\in\conjecturet(\matchinguptot,\agent)$. That is, we require that condition \ref{eq:cc} holds not necessarily for the candidate matchings, which may or may not be a solution, but for the actual solutions (provided they exist). 

Suppose we know the set of \conjectures-solutions is nonempty, but does not satisfy generalized consistency. For concreteness, let \matching\ be a solution for \economyrterminal\ for which a period $t\geq1$ and an agent $\agent\in\calat(\matchinguptot)\cup\calbt(\matchinguptot)$ exist such that $\matchingt(\agent)\neq\agent$ and $\matching\notin\conjecturet(\matchinguptot,\agent)$. Because \matching\ is a solution, the issue of credible blocks discussed above does not arise. However, that \agent\ does not conceive of \matching\ as a potential outcome for the continuation economy suggests a tension exists between what the solution concept prescribes in the continuation economy in which \agent\ is one of the agents available to match, and what \agent\ expects would happen if they would object to \matching. That \matching\ is not part of \agent's conjectures means \agent\ believes \matching\ cannot be enforced in the continuation. Thus, if the solution concept captures the set of self-enforcing outcomes in the continuation economy \emph{and} \agent\ agrees with this assessment,  \matching\ should be part of  \agent's conjectures and the solution concept should satisfy generalized consistency.  As we illustrate below, the rational expectations solution fails both consistency and generalized consistency.


\paragraph{Not all solution concepts have consistent conjectures}
A natural question is how permissive consistency is as a requirement. We illustrate via \autoref{example:re} that the conjectures in the rational expectations solution do not satisfy consistency.\footnote{Because consistency is only a sufficient condition, that the conjectures in the rational expectations solution fail consistency is not enough to conclude the rational expectations solution may be empty.  In recent work, \cite{nicolo2024dynamic} show the rational expectations solution may be empty in dynamic matching markets.} Before introducing the example, defining the rational expectations solution in the context of the dynamic economy is useful; we do so informally here and formally in \autoref{appendix:re}. To that end, fix an economy \economyrterminal\ and a period-1 agent, $\agent\in\arrivala_1\cup\arrivalb_1$. Define the length-\terminal\ economy, $\economyrterminal_{\setminus\agent}$, to be an economy that coincides with \economyrterminal\ except that agent \agent\ arrives in period $2$.\footnote{Formally, if $\agent\in\arrivala_1$, $\economyrterminal_{\setminus\agent}=(\arrivala_1\setminus\{\agent\},\arrivalb_1,\arrivala_2\cup\{\agent\},\arrivalb_2,\dots,\arrivala_\terminal,\arrivalb_\terminal)$.} The definition of the rational expectations solution for \economyrterminal\  builds on the rational expectations solution for $\economyrterminal_{\setminus\agent}$, which we denote by $\re_\terminal(\economyrterminal_{\setminus\agent})$. Under rational expectations, when agent \agent\ considers not being available to match in period $1$, they anticipate that the matching that ensues from period $1$ onward is governed by the solution concept, while respecting \agent's decision to not participate in the period-1 matching. Formally, agent \agent's conjectures are given by  $\re_\terminal(\economyrterminal_{\setminus\agent})$.\footnote{Note the slight abuse of notation. We defined conjectures in \autoref{definition:conjectures} as a subset of $\Matchingsfinal(\emptyset,\agent)$ so that a conjecture specifies that agent \agent\ is unmatched in period $1$. Instead, $\re_\terminal(\economyrterminal_{\setminus\agent})$ is a subset of the set of matchings for $\economyrterminal_{\setminus\agent}$ in which \agent\ only arrives in period $2$ and hence, specifying \agent's outcome in period $1$ is not necessary.} The rational expectations solution for \economyrterminal\ is the set of matchings that satisfy \autoref{definition:recursive} when for all $\agent\in\arrivala_1\cup\arrivalb_1$, conjectures are determined by $\re_\terminal(\economyrterminal_{\setminus\agent})$ and continuation matchings are elements of $\re_{\terminal-1}(\cdot)$.

\begin{example}[Rational expectations does not satisfy consistency]\label{example:re}
Arrivals are given by $\arrivala_1=\{\agenta_1,\agenta_2,\agenta_3\},\arrivalb_1=\{\agentb_1,\agentb_2\},\arrivala_2=\{\agenta_4\}, \arrivalb_2=\{\agentb_3,\agentb_4\}$. Below, we list the agents' preferences. If $(\agentb,0)$ appears before $(\agentbb,1)$ in the ranking of an agent on side \arrivala, they prefer to match immediately with \agentb\ than wait one period to match with \agentbb. That is, the $0$s and $1$s are the exponents of the discount factors, and the list provides the ranking of the discounted utilities $\{\delta_\cdot^{t-1}\payoffa(\cdot,\agentb):\agenta\in\arrivala_1\}$. For the \periodone-agents for whom we present the ranking of utilities alone (i.e., $\agenta_2,\agentb_1$), we assume the discount factor is close to $1$---though not exactly $1$---so that the (static) rankings also represent the rankings over the dynamic matchings.
\begin{align*}
\begin{array}{lccccc}
\agenta_1:&(\agentb_4,0)&(\agentb_2,0)&(\agentb_1,0)&(\agentb_4,1)&\dots\\
\agenta_2:&\agentb_3&\agentb_4&\agentb_2&\agentb_1&\\
\agenta_3:&(\agentb_3,0)&(\agentb_2,0)&(\agentb_3,1)&\dots&\\
\agenta_4:&\agentb_3&\agentb_1&\agentb_4&&
\end{array}
\hspace{1cm}
\begin{array}{lccccc}
\agentb_1:&\agenta_4&\agenta_1&\agenta_2&&\\
\agentb_2:&(\agenta_2,0)&(\agenta_2,1)&(\agenta_3,0)&(\agenta_1,0)&(\agenta_3,1)\\
\agentb_3:&\agenta_3&\agenta_4&\agenta_2&&\\
\agentb_4:&\agenta_4&\agenta_2&\agenta_1&&
\end{array}
\end{align*}
In this economy, the following matching, \matchingstar, is a candidate matching:
\begin{align*}
\matchingstar=\left(\begin{array}{lcl}
\agenta_1&\_\_&\agentb_1\\
\agenta_2&\_\_&\agentb_2\\
\hline
\agenta_3&\_\_&\agentb_3\\
\agenta_4&\_\_&\agentb_4\end{array}\right).
\end{align*}
Furthermore, \matchingstar\ is a rational expectations solution for the economy, which is supported by the following conjectures.\footnote{We provide the supporting details for this example in \autoref{appendix:re}.} Agent $\agenta_2$ cannot block \matchingstar\ because 
\begin{align*}
\matching_{\agenta_2}=\left(\begin{array}{lcl}\agenta_1&\_\_&\agentb_1\\\hline
\agenta_2&\_\_&\agentb_2\\
\agenta_3&\_\_&\agentb_3\\
\agenta_4&\_\_&\agentb_4\end{array}\right),
\end{align*}
is a rational expectations solution in $\economyr^2_{\setminus\{\agenta_2\}}$. Furthermore, agent $\agentb_1$ cannot block because
\begin{align*}
\matching_{\agentb_1}=\left(\begin{array}{lcl}\agenta_3&\_\_&\agentb_2\\\hline\agenta_1&\_\_&\agentb_1\\
\agenta_2&\_\_&\agentb_4\\
\agenta_4&\_\_&\agentb_3\end{array}\right),
\end{align*}
is a rational expectations solution in $\economyr_{\setminus\{\agentb_1\}}^2$. 

However, the conjectures do not satisfy consistency, because \matchingstar\ is not a valid conjecture for $\agenta_3$. In every rational expectations solution in $\economyr_{\setminus\{\agenta_3,\agentb_1\}}^2$, $\agentb_1$ is matched with $\agenta_4$, making  supporting a solution in $\economyr_{\setminus\{\agenta_3\}}^2$, which has $\agentb_1$ matched with $\agenta_1$, impossible. Because \matchingstar\ is a rational expectations solution for the economy, it follows that the solution concept also fails generalized consistency.

The failure of consistency in the rational expectations solution can be tied to the timing assumptions implicit in the solution concept and the conjectures. Given their conjectures, the solution concept presumes the \periodone-agents decide simultaneously whether to be available to match in \periodone. Instead, each agent's conjectures presume they are the first to decide not to be available to match in \periodone, whereas the remaining contemporaneous agents remain available to match in \periodone. For instance, when $\agentb_1$ considers not being available to match in \periodone, $\agentb_1$ presumes $\agenta_3$ is available to match in \periodone, and in fact, $\agenta_3$ must match in \periodone\ to prevent $\agentb_1$ from blocking \matchingstar. Instead, $\agenta_3$'s conjectures presume $\agenta_3$ cannot match in \periodone, and thus, $\agentb_1$ is able to match with $\agenta_4$ in \periodtwo. For this reason, \matchingstar\ cannot be compatible with $\agenta_3$'s conjectures: supporting \matchingstar\ as a rational expectations solution requires that $\agenta_3$ may possibly match in \periodone.
\end{example}

The analysis has so far identified a property of the family of conjectures, \conjectures, that guarantees existence. Yet, we have said little about what these conjectures entail. For instance, the conjectures could include continuation matchings that are, in some sense, not credible; a property that would make the \conjectures-solutions less appealing. Conjectures such as those in \autoref{eq:eqbm-cont} satisfy consistency and avoid these ``non-credible" threats. However, they impose no structure on the matching among the period-1 agents. \autoref{sec:solutions} introduces two families of conjectures (and their corresponding solution concepts) addressing these issues.

%

\section{Two solution concepts with consistent conjectures}\label{sec:solutions}
In this section, we introduce two families of consistent conjectures and their corresponding solution concepts. \autoref{sec:cvr} defines \solone, a refinement of dynamic stability in \cite{doval2022dynamically}. \autoref{sec:sds} defines sophisticated dynamic stability, an extension to dynamic matching markets of the eponymous solution concept in \cite{hafalir2008stability}.

\subsection{Continuation-values-respecting dynamic stability}\label{sec:cvr}
The solution concept we introduce in this section, \solone, builds on the conjectures in \autoref{eq:eqbm-cont}. Whereas the conjectures in \autoref{eq:eqbm-cont} require agents to believe the continuation matching abides by the solution concept, they place no restriction on what period-$t$ matching an agent expects would ensue conditional on not matching in period $t$. \solone\ imposes a minimal requirement on the period-$t$ matching that reflects other agents' continuation values. In what follows, we first define the requirement on the period-$t$ matching (\autoref{definition:satwm}), the conjectures for \solone\ (see \autoref{eq:solonec}), and the solution concept (\autoref{definition:cvr-dynamic}). We then show in \autoref{proposition:cvr} that \solone\ has consistent conjectures. We end the section by discussing the connection with dynamic stability in \cite{doval2022dynamically}.

To define \solone, we first introduce a condition on the period-$t$ matching. To that end, fix an economy, \economyrterminal, and a matching through period $t-1$, \matchinguptot.  Suppose that for each agent $\agent\in\calat(\matchinguptot)\cup\calbt(\matchinguptot)$, we are given a set of matchings $\conjectureb_\agent\subset\Matchingsfinal(\matchinguptot,\agent)$. Similar to \autoref{definition:one-shot-economy}, we can define the one-period economy with agents in $\calat(\matchinguptot)\cup\calbt(\matchinguptot)$ induced by $\{\conjectureb_\agent:\agent\in\calat(\matchinguptot)\cup\calbt(\matchinguptot)\}$, which we denote by $\economyr_{t,\conjectureb}$. Given a period-$t$ matching, \matchingt, let $\economyr_{t,\conjectureb}^{\matchingt}$ denote the restriction of $\economyr_{t,\conjectureb}$ to the set of agents $\agent\in\calat(\matchinguptot)\cup\calbt(\matchinguptot)$ such that $\matchingt(\agent)\neq\agent$.

\begin{definition}[Stability among those who match relative to continuation values]\label{definition:satwm} The period-$t$ matching, \matchingt, is \emph{stable among those who match in period $t$ relative to $\conjectureb$} if the restriction of \matchingt\ to $\economyr_{t,\conjecture}^{\matchingt}$ is an element of $\stable(\economyr_{t,\conjecture}^{\matchingt})$, that is, $\matchingt|_{\economyr_{t,\conjecture}^{\matchingt}}\in\stable(\economyr_{t,\conjecture}^{\matchingt})$.
\end{definition}
As the name suggests, \autoref{definition:satwm} requires that the period-$t$ matching satisfies a condition analogous to the static notion of stability among the agents who exit the economy in period $t$. We say ``analogous'' because rather than requiring that the period-$t$ matching be individually rational, \autoref{definition:satwm} requires that each agent \agent\ be matched to a partner they prefer to the worst matching in $\conjecture_\agent$.\footnote{
If for each agent $\agent$, we take $\conjectureb_\agent^{IR}$ to be the subset of individually rational matchings for \agent\ in $\Matchingsfinal(\matchinguptot,\agent)$, we obtain the condition in \citet[Definition 4]{doval2022dynamically}. We discuss the connection with \cite{doval2022dynamically} in detail below.}

Given \autoref{definition:satwm}, we now define the conjectures associated with \solone. 	Suppose we have already defined \solone\ for economies of length $\{1,\dots,\terminal-t\}$ and let $\dsh_{\terminal-t}$ denote the correspondence of  \cvr\ dynamically stable matchings for length $\terminal-t$ economies. Fix agent $\agent\in\calat(\matchinguptot)\cup\calbt(\matchinguptot)$ and define their conjectures under \solone\ as follows:
\begin{align}\label{eq:solonec}
\solonect(\matchinguptot,\agent)=\left\{\matchingb\in\Matchingsfinal(\matchinguptot,\agent):\begin{array}{ll}\text{ (i) }\matchingbt|_{\economyr_{t,\solonect}^{\matchingbt}}\in\stable\left(\economyr_{t,\solonect}^{\matchingbt}\right)\\\text{ (ii) }(\matchingb_s)_{s=t+1}^\terminal|_{\economyrterminal_{t+1}(\matchingbuptot)}\in\dsh_{\terminal-t}(\economyrterminal_{t+1}(\matchingbuptot))\end{array}\right\}.
\end{align}
That is, when considering not being available to match in period $t$, \agent's conjectured matchings satisfy two properties. First, \agent\ assumes the continuation matching will be a \cvr\ dynamically stable matching. Second, \agent\ conjectures that those agents who exit the economy in period $t$ under \matchingbt\ could not have found a better matching among themselves. That is, \matchingbt\ cannot be improved by pairwise blocks, and no agent who matches in period $t$ matches with someone worse than their implicit continuation value according to their own conjectures, $\solonect$.\footnote{The last part is consistent with the idea that agents simultaneously decide whether to match or remain unmatched given their set of conjectures (recall the discussion in \autoref{example:re}).} This  last aspect of \agent's conjectures is the sense in which they respect other contemporaneous agents' continuation values (as defined by their conjectures).  Despite the fixed-point flavor of the set in \autoref{eq:solonec}, agent \agent's conjectures are well defined: because \agent\ is unmatched in period $t$ under \matchingb, the set $\stable\left(\economyr_{t,\solonect}^{\matchingbt}\right)$ can be defined independently of agent \agent's conjectures. 

%
%
%

We can define the set of \cvr\ dynamically stable matchings for $\economyrterminal_t(\matchinguptot)$, $\dsh_{\terminal-(t-1)}(\economyrterminal_t(\matchinguptot))$, to be the set of matchings for this economy that satisfy \autoref{definition:recursive} with $\solonect(\matchinguptot,\cdot)$ as the conjectures and $\dsh_{\terminal-t}$ as the continuation solution concept. Working recursively in this way, we can define the family of conjectures $\conjectures_{\dsh}$ for \economyrterminal\ to be the collection $\{\solonect(\matchinguptot,\agent):t\geq1,\matchinguptot,\agent\in\calat(\matchinguptot)\cup\calbt(\matchinguptot)\}$. The definition of \solone\ is now immediate:
\begin{definition}[\solone]\label{definition:cvr-dynamic}
A matching \matching\ for \economyrterminal\ is \cvr\ dynamically stable if it is a $\conjectures_{\dsh}$-solution for \economyrterminal. We denote by $\dsh_\terminal(\economyrterminal)$ the set of such solutions.
\end{definition}

\autoref{proposition:cvr} is the main result of this section:
\begin{proposition}\label{proposition:cvr} The family $\conjectures_{\dsh}$ satisfies consistency. Consequently, for all $\terminal\in\naturals$, the correspondence $\dsh_\terminal$ is nonempty valued.
\end{proposition}
The proof of \autoref{proposition:cvr} is in \autoref{appendix:solutions}. The proof proceeds by induction on the length of the economy, first showing the set of conjectures is nonempty and then showing they satisfy consistency; \autoref{theorem:consistent-conjectures} then implies the set of \cvr\ dynamically stable matchings is nonempty. Thus, the key step is showing the conjectures are nonempty: Because of the fixed-point character of the conjectures in period $t$, we build a recursion whose iteration delivers  the set \solonect\ as a limit.

\paragraph{Comparison with dynamic stability \citep{doval2022dynamically}} \solone\ is a refinement of dynamic stability, which is the stability notion in \cite{doval2022dynamically}. In particular, if we denote by \dsc\ the conjectures corresponding to dynamic stability, one can show that for all $t$, all $\matchinguptot$, and $\agent\in\calat(\matchinguptot)\cup\calbt(\matchinguptot)$, the conjectures $\solonect(\matchinguptot,\agent)\subseteq\conjecture_{\ds,t}(\matchinguptot,\agent)$, which in turn implies the set of \cvr\ dynamically stable matchings is a subset of those that are dynamically stable. 

To understand the difference between the solution concepts, comparing the conjectures in \autoref{eq:solonec} with those in \cite{doval2022dynamically} is instructive. Whereas both solution concepts require agents to conjecture that the continuation matching conforms with the solution concept in the induced continuation economy (i.e., requirement (ii) in \autoref{eq:solonec}), the conjectures in \cite{doval2022dynamically} impose a weaker requirement on the period-$t$ matching, namely, that among the agents who exit the economy in period $t$, the  period-$t$ matching is individually rational and no blocking pairs exist.\footnote{Formally, let $\conjecture_\agent^{IR}$ denote the set of matchings in $\Matchingsfinal(\matchinguptot,\agent)$ that are individually rational for \agent. Then, in dynamic stability, matching $\matchingb\in\Matchingsfinal(\matchinguptot,\agent)$ is a conjecture for \agent\ if (i) \matchingbt\ is stable among those who match relative to continuation values $\conjecture_\cdot^{IR}$ and (ii) the continuation matching is dynamically stable.}  In other words, under dynamic stability, an agent \agent\ conjectures that other contemporaneous agents  are willing to exit in period $t$ with partners that dominate being unmatched through period \terminal, even if they may not dominate the worst conjectured matching under $\dsct(\matchinguptot,\cdot)$. At the same time, when considering whether to block the proposed matching in period $t$, the agents available to match in period $t$ compare their payoff under the proposed matching with their worst conjectured matching under $\dsct(\matchinguptot,\cdot)$. That is, in dynamic stability, agents make their decisions to be available to match in period $t$ relative to their conjectures, but these conjectures do not take into account that other contemporaneous agents use their own conjectures when evaluating whether to match in period $t$. \solone\ eliminates this gap by requiring the period-$t$ agents to internalize that other contemporaneous agents' continuation values are exactly those prescribed by their conjectures.


We conclude this section illustrating that \solone\ is a strict refinement of dynamic stability:
\begin{example}[$\dsh\subsetneq\ds$]\label{example:dsh-neq-ds} This example is due to Antonio Nicol\`o, Pietro Salmaso, and Riccardo Saulle. Arrivals are given by $\arrivala_1=\{\agenta_1,\agenta_2\},\arrivalb_1=\{\agentb_1,\agentb_2\},\arrivala_2=\{\agenta_3,\agenta_4\},\arrivalb_2=\{\agentb_3,\agentb_4\}$. Assume the discount factors of the period-1 agents are close to $1$, so that the (static) rankings below are enough to describe the agents' preferences over dynamic matchings.
\begin{align*}
\begin{array}{lllll}
\agenta_1:&\agentb_3&\agentb_4&\agentb_1&\\
\agenta_2:&\agentb_4&\agentb_2&\agentb_3&\\
\agenta_3:&\agentb_1&\agentb_4&\agentb_2&\\
\agenta_4:&\agentb_1&\agentb_4&\agentb_2&\agentb_3
\end{array}
\qquad\qquad\qquad
\begin{array}{lllll}
\agentb_1:&\agenta_3&\agenta_4&\agenta_1&\\
\agentb_2:&\agenta_2&\agenta_3&\agenta_4&\\
\agentb_3:&\agenta_1&\agenta_2&\agenta_4&\\
\agentb_4:&\agenta_1&\agenta_3&\agenta_2&\agenta_4
\end{array}
\end{align*}
Consider the following three matchings:
\begin{align*}
\matching^L=\left(\begin{array}{lcl}\agenta_2&\_&\agentb_2\\\hline\agenta_1&\_&\agentb_3\\\agenta_3&\_&\agentb_1\\\agenta_4&\_&\agentb_4\end{array}\right)
\quad\quad\matching^C=\left(\begin{array}{lcl}\agenta_1&\_&\agentb_1\\\hline\agenta_2&\_&\agentb_2\\\agenta_3&\_&\agentb_4\\\agenta_4&\_&\agentb_3\end{array}\right)\quad\quad
\matching^R=\left(\begin{array}{lcl}&\emptyset&\\\hline\agenta_1&\_&\agentb_3\\\agenta_2&\_&\agentb_4\\\agenta_3&\_&\agentb_1\\\agenta_4&\_&\agentb_2\end{array}\right).
\end{align*}
The matching on the left, $\matching^L$, is a dynamically stable matching, with $\matching^C$ describing the conjectured matching that dissuades $\agenta_2$ from waiting to be matched in \periodtwo. Importantly, under $\matching^C$, $\agenta_1$ and $\agentb_1$ match in period $1$. However, we can easily see $\agenta_1$ can always guarantee matching with $\agentb_3$ (and similarly, $\agentb_1$  with $\agenta_3$) by remaining unmatched in \periodone. In other words, $\matching^C$ cannot be a conjecture for $\agenta_2$ under \solonec, and hence, $\matching^L$ is not \cvr\ dynamically stable. Instead, $\matching^R$ is \cvr\ dynamically stable (and hence, dynamically stable).
\end{example}


\subsection{Sophisticated dynamic stability}\label{sec:sds}
In this section, we provide the natural extension of the solution concept in \cite{hafalir2008stability} to the dynamic environment we study. Because \cite{hafalir2008stability} names the solution concept \emph{sophisticated expectations}, we denote ours by \soltwo. Like \solone, sophisticated dynamic stability is defined recursively. In what follows, we denote by $\sds_t$ the correspondence that maps economies of length $t$ to their set of sophisticated dynamically stable matchings. 

Agents' conjectures in sophisticated dynamic stability are defined recursively, building on two elements. First, like the rational expectations solution, conjectures in \soltwo\ satisfy a form of coherence in the economy induced by an agent's decision to not be available to match. Recall that when agent \agent\ chooses to remain unmatched in period $t$, it is \emph{as if} \agent\ induces a new economy from period $t$ onward, where everything is as before, except that now \agent\ arrives in period $t+1$. Similar to the rational expectations solution, \soltwo\ requires that \agent's conjectures include all matchings \matchingb\ that from period $t$ onward are sophisticated dynamically stable in the economy induced by \agent\ remaining unmatched in period $t$ (see \autoref{eq:re} below).  We know from \autoref{example:re} that this property alone does not deliver a solution concept with consistent conjectures. The second element is a procedure that guarantees the consistency of the conjectures in \soltwo, by recursively expanding the set of conjectures to guarantee their consistency.

We now define the conjectures for \soltwo, which we denote by $\conjecture_{\sds,t}$. Fix an economy \economyrterminal\ and a matching through period $t-1$, \matchinguptot. Recall \matchinguptot\ induces a \emph{continuation} economy of length $\terminal-(t-1)$, $\economyrterminal_t(\matchinguptot)$. Suppose that for all $\agent\in\calat(\matchinguptot)\cup\calbt(\matchinguptot)$, we have defined the set of sophisticated dynamically stable matchings for the economy of length $T-(t-1)$ induced by \agent\ arriving in period $t+1$, $\sds_{\terminal-(t-1)}\left(\economyr_t^\terminal(\matchinguptot)_{\setminus\agent}\right)$. 

 For each $\agent\in\calat(\matchinguptot)\cup\calbt(\matchinguptot)$, the conjectures in \soltwo\ are defined recursively, starting from the following set:
 \small
\begin{align}\label{eq:re}
\addon_0(\economyrterminal_t(\matchinguptot),\agent)=\left\{\matchingb\in\Matchingsfinal(\matchinguptot,\agent):(\matchingb_t\setminus\agent,(\matchingb_s)_{s=t+1}^{T})|_{\economyr_t^\terminal(\matchinguptot)_{\setminus\agent}}\in\sds_{\terminal-(t-1)}(\economyr_t^\terminal(\matchinguptot)_{\setminus\agent})\right\},
\end{align}\normalsize
where (i) $\matchingb_t\setminus\agent$ is a period-$t$ matching on $\arrivalauptot\cup\arrivalbuptot\setminus\{\agent\}$ that coincides with \matchingbt\ on that domain, and (ii) we make the dependence of $\addon_0$ on the continuation economy explicit, which is useful because of the recursive nature of the definition.\footnote{The set of sophisticated dynamically stable matchings for $\economyrterminal_t(\matchinguptot)$ depends on the set of sophisticated dynamically stable matchings for $\economyrterminal_t(\matchinguptot)_{\setminus\agent}$, which in turn depends on the conjectures $\conjecture_{\sds,t}(\economyrterminal_t(\matchinguptot)_{\setminus\agent},\agent^\prime)$ for $\agent^\prime\in\calat(\matchinguptot)\cup\calbt(\matchinguptot)\setminus\{\agent\}$, and so on.}

In words, the set $\addon_0(\economyrterminal_t(\matchinguptot),\agent)$ consists of all sophisticated dynamically stable matchings for the $\terminal-(t-1)$ economy in which \agent\ arrives in period $t+1$. As the construction below makes clear, $\addon_0(\economyrterminal_t(\matchinguptot),\agent)$ is always a subset of \agent's conjectures under \soltwo, and hence, \soltwo\ satisfies a form of coherence similar to that of the rational expectations solution. In fact, if we defined the conjectures of \soltwo\ using \autoref{eq:re}, \soltwo\ would coincide with the rational expectations solution. 

Given the collection $\{\addon_0(\economyrterminal_t(\matchinguptot),\agent):\agent\in\calat(\matchinguptot)\cup\calbt(\matchinguptot)\}$, we can now construct the set of candidate matchings given ``conjectures'' $\addon_0(\cdot)$ \emph{and} continuation matchings taken from $\sds_{\terminal-t}$:
\begin{align*}
\candidate\left(\matchinguptot,\addon_0,\sds_{\terminal-t}\right)=\left\{\matchingb\in\Matchingsfinal(\matchinguptot):\begin{array}{l}\text{(i) }\matchingb_t|_{\economyr_{t,\addon_0}}\in\stable(\economyr_{t,\addon_0})\\
    \text{(ii) }(\matchingb_s)_{s=t+1}^T|_{\economyr_{t+1}^\terminal(\matchingb^t)}\in\sds_{\terminal-t}(\economyr_{t+1}^\terminal(\matchingb^t))\end{array}\right\}.
\end{align*}
Suppose now a matching $\matchingb\in\candidate\left(\matchinguptot,\addon_0,\sds_{\terminal-t}\right)$ and an agent $\agent\in\calat(\matchinguptot)\cup\calbt(\matchinguptot)$ exist such that $\matchingbt(\agent)=\agent$ and $\matchingb\notin\addon_0(\economyrterminal_t(\matchinguptot),\agent)$. In other words, \matchingb\ fails property \ref{eq:cc} when using the sets $\addon_0$ as conjectures. Then, we expand the set $\addon_0$ by adding these missing ``candidate'' matchings, that is,
\begin{align}\label{eq:sds-1}
    \addon_1(\economyrterminal_t(\matchinguptot),\agent)=\addon_{0}(\economyrterminal_t(\matchinguptot),\agent)\cup\left\{\matchingb\in\candidate(\matchinguptot,\addon_{0},\sds_{\terminal-1}):\matchingb_t(k)=k\right\}.
\end{align}
Adding these additional matchings to $\addon_0$ is meant to solve the issue discussed in \autoref{sec:consistent-conjectures}: Whenever consistency fails and a ``candidate'' matching,  $\matchingb\in\candidate\left(\matchinguptot,\addon_0,\sds_{\terminal-t}\right)$, fails to be a solution, it is blocked by a pair of agents, at least one of which prefers their worst conjectured matching in $\addon_0$ to matching together. By adding the matchings in $\candidate(\matchinguptot,\addon_0,\sds_{\terminal-1})$, we ensure agents do not dismiss matchings such as \matchingb, whose blocks are not credible.

Because $\addon_1$ includes $\addon_0$, the set of candidate matchings given ``conjectures'' $\addon_1$, $\candidate(\matchinguptot,\addon_1,\sds_{\terminal-1})$, may include new ``candidate'' matchings not in $\candidate(\matchinguptot,\addon_0,\sds_{\terminal-1})$. Furthermore, these new matchings may also be ruled out by blocks that are not credible given ``conjectures'' $\addon_1$. To address this possibility, we continue to expand the ``conjectures'' recursively as follows: for $n\geq 1$ and $\agent\in\calat(\matchinguptot)\cup\calbt(\matchinguptot)$, define
\begin{align}\label{eq:recursion}
    \addon_n(\economyrterminal_t(\matchinguptot),\agent)=\addon_{n-1}(\economyrterminal_t(\matchinguptot),\agent)\cup\left\{\matchingb\in\candidate(\matchinguptot,\addon_{n-1},\sds_{\terminal-1}):\matchingb_t(k)=k\right\}.
\end{align}
This process expands the sets $\addon_{n-1}$ of those agents who find themselves unmatched in the economy with continuation values determined by $\addon_{n-1}(\economyrterminal_t(\matchinguptot),\cdot)$. Note matchings are only added if the elements in $\candidate(\matchinguptot,\addon_{n-1},\sds_{\terminal-1})$ are not part of the set $\addon_{n-1}$. 

We define \agent's conjectures as the limit of the set $\addon_n(\economyrterminal_t(\matchinguptot),\agent)$; that is,
\begin{align}\label{eq:rs-conjectures}
\conjecture_{\sds,t}(\economyrterminal_t(\matchinguptot),\agent)=\lim_{n\rightarrow\infty}\addon_n(\economyrterminal_t(\matchinguptot),\agent).
\end{align}
The set in \autoref{eq:rs-conjectures} is well defined because the sequence $\addon_n$ is increasing (in the set-inclusion sense) and bounded above by $\Matchingsfinal(\matchinguptot,\agent)$. By definition, each element of $\candidate(\matchinguptot,\conjecture_{\sds,t},\sds_{\terminal-t})$ satisfies property \ref{eq:cc} when using the sets $\conjecture_{\sds,t}$ as conjectures, so the conjectures do not continue to expand once $\conjecture_{\sds,t}$ is reached. 

We use \autoref{example:re} to illustrate the construction of $\sdsc$:
\setcounter{example}{0}
\begin{example}[continued]\label{example:re-c}
Recall that in the economy of \autoref{example:re} the conjectures in the rational expectations solution fail consistency because the candidate matching, denoted by \matchingstar\ in the example, is not part of agent $\agenta_3$'s conjectures. For that reason, we focus in what follows on describing $\agenta_3$'s conjectures under \soltwo. 

In this example, for each period-$1$ agent, the set $F_0(\economyrtwo,\agent)$ corresponds to their conjectures under rational expectations. The reason is that, for each $\agent\in\arrivala_1\cup\arrivalb_1$, the set of sophisticated dynamically stable matchings and the set of rational expectations solutions for $\economyrtwo_{\setminus\agent}$ coincide. In particular, in the economy induced by agent $\agenta_3$ remaining unmatched in period $1$, $\economyrtwo_{\setminus\agenta_3}$, the following matching is the unique sophisticated dynamically stable matching:
\begin{align*}
\matching_{\agenta_3}=\left(\begin{array}{lcl}\agenta_1&\_\_&\agentb_2\\\hline
\agenta_2&\_\_&\agentb_4\\
\agenta_3&\_\_&\agentb_3\\
\agenta_4&\_\_&\agentb_1\end{array}\right),
\end{align*}
and hence, $F_0(\economyrtwo,\agenta_3)=\{\matching_{\agenta_3}\}$. 

The same arguments as those in \autoref{sec:consistent-conjectures} imply the set $\candidate(\emptyset,F_0,\sds_1)$ is a singleton, consisting of \matchingstar. Note agent $\agenta_3$ is the only period-$1$ agent who is unmatched under \matchingstar\ and $\matchingstar\notin\addon_0(\economyrtwo,\agenta_3)$. Thus, $\addon_1(\economyrtwo,\agenta_3)=\{\matching_{\agenta_3},\matchingstar\}$, whereas for $\agent\neq\agenta_3$, $\addon_1(\economyrtwo,\agent)=\addon_0(\economyrtwo,\agent)$. Because $\agenta_3$ is matched with $\agentb_3$ under both \matchingstar\ and $\matching_{\agenta_3}$, the set of candidate matchings given $\addon_1$, $\candidate(\emptyset,\addon_1,\sds_1)$, equals $\candidate(\emptyset,\addon_0,\sds_1)=\{\matchingstar\}$. Because $\matchingstar\in\addon_1(\economyrtwo,\agenta_3)$, no matching is added to $\agenta_3$'s ``conjectures'' at this step. We then conclude $\conjecture_{\sds,1}(\economyrtwo,\agenta_3)=\{\matchingstar,\matching_{\agenta_3}\}$. 
\end{example}

Having defined the conjectures under \soltwo\ for $\economyrterminal_t(\matchinguptot)$, we define its set of sophisticated dynamically stable matchings, $\sds_{\terminal-(t-1)}(\economyrterminal_t(\matchinguptot))$, to be the set of matchings for this economy that satisfy \autoref{definition:recursive} with $\conjecture_{\sds,t}(\economyrterminal_t(\matchinguptot),\cdot)$ as the conjectures and $\sds_{\terminal-t}$ as the continuation solution concept.\footnote{To be precise, for a given subset $\coalition\subseteq\calat(\matchinguptot)\cup\calbt(\matchinguptot)$, 
we should define the set of sophisticated dynamically stable matchings for $\economyrterminal_{t}(\matchinguptot)_{\setminus\coalition}$, having defined $\sds_{\terminal-(t-1)}(\economyrterminal_t(\matchinguptot)_{\setminus\coalition^\prime})$ for all subsets $\coalition^\prime\subseteq(\calat(\matchinguptot)\cup\calbt(\matchinguptot))\setminus\coalition$, where $\economyrterminal_{t}(\matchinguptot)_{\setminus\coalition}$ is the length-$\terminal-(t-1)$ economy in which the agents in \coalition\ arrive in period $t+1$. However, $\economyrterminal_t(\matchinguptot)_{\setminus\coalition}$ is yet another economy of length $\terminal-(t-1)$, so this omission is hopefully innocuous.} Working recursively, we can define the family of conjectures $\conjectures_{\sds}$ for \economyrterminal. The definition of \soltwo\ is now immediate:
\begin{definition}[\Soltwo]\label{definition:sds}
A matching \matching\ for \economyrterminal\ is sophisticated dynamically stable if it is a $\conjectures_{\sds}$-solution for \economyrterminal. We denote by $\sds_{\terminal}(\economyrterminal)$ the set of such solutions.
\end{definition}
When applied to the economy of \autoref{example:re-c},  \autoref{definition:sds} implies \matchingstar\ is the unique sophisticated dynamically stable matching. That is, in this example, \soltwo\ coincides with the rational expectations solution, but different conjectures support \matchingstar\ as a solution in each case.

Because the conjectures in \soltwo\ are constructed so that they satisfy consistency, the following result is an immediate corollary of \autoref{theorem:consistent-conjectures}:
\begin{corollary}
For all $\terminal\in\naturals$, $\sds_\terminal$ is a nonempty-valued correspondence.
\end{corollary}

\paragraph{Comparison with dynamic stability and \cvr-dynamic stability} We conclude this section by comparing \soltwo\ with the solution concepts discussed in \autoref{sec:cvr}.

Our first observation is summarized in  \autoref{proposition:sds-refines} below: sophisticated dynamic stability is a refinement of dynamic stability:
\begin{proposition}\label{proposition:sds-refines}
For all $\terminal\in\naturals$, $\sds_\terminal\subseteq\ds_\terminal$. 
\end{proposition}
The proof of \autoref{proposition:sds-refines} is in \autoref{appendix:solutions}. We establish that the conjectures in sophisticated dynamic stability are a subset of those under dynamic stability. Therefore, a sophisticated dynamically stable matching is dynamically stable, but as \autoref{example:cvr-not-sds} below shows, the opposite may not be true. Because, in this example, the set of \cvr\ dynamically stable matchings coincides with those that are dynamically stable, \autoref{example:cvr-not-sds} also illustrates that in some economies, \soltwo\ is a refinement of \solone. 
\begin{example}[$\sds\subsetneq\ds$ and $\cvr\neq\sds$]\label{example:cvr-not-sds}
Consider the following economy. Arrivals are $\arrivala_1=\{\agenta_1,\agenta_2,\agenta_3\}$, $B_1=\{\agentb_1\}$, $\arrivala_2=\emptyset$, $B_2=\{\agentb_2,\agentb_3\}$. Preferences are given by:
\[\begin{array}{ccccccc}
\agenta_1:&(\agentb_2,0)&(\agentb_2,1)&(\agentb_1,0)&(\agentb_1,1)&(\agentb_3,0)&(\agentb_3,1)\\
\agenta_2:&(\agentb_2,0)&(\agentb_2,1)&(\agentb_3,0)&(\agentb_1,0)&(\agentb_3,1)&(\agentb_1,1)\\
\agenta_3:&(\agentb_3,0)&(\agentb_3,1)&(\agentb_2,0)&(\agentb_2,1)&(\agentb_1,0)&(\agentb_1,1)
\end{array}\hspace{0.5cm}
\begin{array}{ccccccc}
\agentb_1:&(\agenta_1,0)&(\agenta_2,0)&(\agenta_1,1)&&&\\
\agentb_2:&\agenta_3&\agenta_1&\agenta_2&&&\\
\agentb_3:&\agenta_2&\agenta_3&&&&
\end{array}
\]
The following three matchings are \cvr\ dynamically stable, and hence, dynamically stable:
\[\matching^L=\left(\begin{array}{lcl}
\agenta_2&\_\_&\agentb_1\\
\hline
\agenta_1&\_\_&\agentb_2\\
\agenta_3&\_\_&\agentb_3
\end{array}\right)
\hspace{1cm}
\matching^C=\left(\begin{array}{lcl}
\agenta_1&\_\_&\agentb_1\\
\hline
\agenta_2&\_\_&\agentb_3\\
\agenta_3&\_\_&\agentb_2
\end{array}\right)
\hspace{1cm}
\matching^R=\left(\begin{array}{lcl}
\agenta_1&\_\_&\agentb_1\\
\hline
\agenta_2&\_\_&\agentb_2\\
\agenta_3&\_\_&\agentb_3
\end{array}\right).
\]
The reason is that the following matching is an element of $\conjecture_{\ds,1}(\emptyset,\agent)$ and of $\conjecture_{\dsh,1}(\emptyset,\agent)$ for $\agent\in\{\agenta_1,\agenta_2,\agentb_1\}$:
\[\matching_\emptyset=\left(\begin{array}{lcl}
&\emptyset&\\
\hline
\agenta_1&\_\_&\agentb_1\\
\agenta_2&\_\_&\agentb_3\\
\agenta_3&\_\_&\agentb_2
\end{array}\right).\]
Because no one matches in \periodone\ under $\matching_\emptyset$, the additional conditions that \solone\ imposes on the conjectures do not rule out matching $\matching_\emptyset$ as an element of the conjectures under \solone.

Instead, only $\matching^L$ is sophisticated dynamically stable. As we show in \autoref{appendix:solutions}, $\matching^\emptyset$ is not part of the conjectures of $\agenta_1$, because $(\agenta_2,\agentb_1)$ block $\matching_\emptyset$   in the two-period economy induced by $\agenta_1$ waiting to be matched, $\economyrtwo_{\setminus\{\agenta_1\}}=(\arrivala_1\setminus\{\agenta_1\},\arrivalb_1,\{\agenta_1\},\arrivalb_2)$. In fact,  the set $\conjecture_{\sds,1}(\emptyset,\agenta_1)$ is a singleton consisting of $\matching^L$, implying that by waiting to be matched, $\agenta_1$ can guarantee the payoff of matching with $\agentb_2$. 
\end{example}
Although \autoref{example:cvr-not-sds} shows that in some economies, \soltwo\ refines \solone, whether this finding holds more generally is an open question. 

\section{Conclusions}\label{sec:conclusions}
In this paper, we provide a unifying framework to study existence properties of solution concepts for two-sided, one-to-one, dynamic matching markets in which matching is irreversible. Following the literature on matching with externalities \citep{sasaki1996two,hafalir2008stability}, we give center stage to agents' conjectures about the matching that would ensue should they remain unmatched. Given a collection of conjectures, we define a solution concept given these conjectures. We identify a sufficient condition on the conjectures for the corresponding solution concept to be nonempty (\autoref{theorem:consistent-conjectures}). Armed with this result, we propose two families of conjectures and, consequently, two new solution concepts for dynamic matching markets with nonempty solutions.

We see several avenues worth exploring and left for future work. First, whereas the rational expectations solution does not satisfy consistency, \cite{nicolo2023dynamic} show its existence is guaranteed in dynamic, one-to-one, one-sided matching markets. Because the same externality is present in both two-sided and one-sided dynamic matching markets, understanding what sets apart these two environments is useful.

Second, whereas we propose two new solution concepts for dynamic matching markets, they are by no means the only solution concepts that satisfy consistency. Further understanding the set of solution concepts that satisfy consistency would allow us to delineate the limits of the predictions afforded by such solution concepts. \cite{nicolo2024dynamic} is a step in this direction. Relatedly, whereas consistent conjectures ensure the existence of \conjectures-solutions, whether they are necessary for existence is worth exploring. 

Finally, because important dynamic matching applications, such as sequential assignment in school choice, involve many-to-one matching, extending our results to many-to-one matching markets is natural. As \cite{altinok2019dynamic} and \cite{liu2023stability} highlight, new issues arise in the study of such markets, and understanding what the analogue of consistency is in these markets is worth exploring.


\bibliographystyle{ecta}
\bibliography{matching}

\begin{thebibliography}{23}
\newcommand{\enquote}[1]{``#1''}
\expandafter\ifx\csname natexlab\endcsname\relax\def\natexlab#1{#1}\fi

\bibitem[\protect\citeauthoryear{Altinok}{Altinok}{2019}]{altinok2019dynamic}
\textsc{Altinok, A.} (2019): \enquote{Dynamic Many-to-One Matching,}
  \emph{Available at SSRN 3526522}.

\bibitem[\protect\citeauthoryear{Chowdhury}{Chowdhury}{2004}]{chowdhury2004marriage}
\textsc{Chowdhury, P.~R.} (2004): \enquote{Marriage markets with
  externalities,} Tech. rep., Indian Statistical Institute, New Delhi, India.

\bibitem[\protect\citeauthoryear{Chwe}{Chwe}{1994}]{chwe1994farsighted}
\textsc{Chwe, M.} (1994): \enquote{Farsighted coalitional stability,}
  \emph{Journal of Economic theory}, 63, 299--325.

\bibitem[\protect\citeauthoryear{Doval}{Doval}{2015}]{doval2015theory}
\textsc{Doval, L.} (2015): \enquote{A Theory of Stability in Dynamic Matching
  Markets,}
  \href{http://economics.yale.edu/sites/default/files/doval_jmp.pdf}{Click
  here.}

\bibitem[\protect\citeauthoryear{Doval}{Doval}{2021}]{doval2020dynamically}
---\hspace{-.1pt}---\hspace{-.1pt}--- (2021): \enquote{Dynamically Stable
  Matching,} \emph{arXiv preprint arXiv:1906.11391}.

\bibitem[\protect\citeauthoryear{Doval}{Doval}{2022}]{doval2022dynamically}
---\hspace{-.1pt}---\hspace{-.1pt}--- (2022): \enquote{Dynamically stable
  matching,} \emph{Theoretical Economics}, 17, 687--724.

\bibitem[\protect\citeauthoryear{Doval}{Doval}{Forthcoming}]{doval2023dynamic}
---\hspace{-.1pt}---\hspace{-.1pt}--- (Forthcoming): \enquote{Dynamic
  Matching,} in \emph{Handbook of the Economics of Matching}, ed. by Y.-K. Che,
  P.-A. Chiappori, and B.~Salanie, Elsevier.

\bibitem[\protect\citeauthoryear{Gale and Shapley}{Gale and
  Shapley}{1962}]{gale1962college}
\textsc{Gale, D. and L.~S. Shapley} (1962): \enquote{College admissions and the
  stability of marriage,} \emph{American Mathematical Monthly}, 9--15.

\bibitem[\protect\citeauthoryear{Hafalir}{Hafalir}{2008}]{hafalir2008stability}
\textsc{Hafalir, I.~E.} (2008): \enquote{Stability of marriage with
  externalities,} \emph{International Journal of Game Theory}, 37, 353--369.

\bibitem[\protect\citeauthoryear{Harsanyi}{Harsanyi}{1974}]{harsanyi1974equilibrium}
\textsc{Harsanyi, J.~C.} (1974): \enquote{An equilibrium-point interpretation
  of stable sets and a proposed alternative definition,} \emph{Management
  science}, 20, 1472--1495.

\bibitem[\protect\citeauthoryear{Kadam and Kotowski}{Kadam and
  Kotowski}{2018}]{kadam2018multiperiod}
\textsc{Kadam, S.~V. and M.~H. Kotowski} (2018): \enquote{Multiperiod
  Matching,} \emph{International Economic Review}, 59, 1927--1947.

\bibitem[\protect\citeauthoryear{Kimya}{Kimya}{2022}]{kimya2022farsighted}
\textsc{Kimya, M.} (2022): \enquote{Farsighted objections and maximality in
  one-to-one matching problems,} \emph{Journal of Economic Theory}, 204,
  105499.

\bibitem[\protect\citeauthoryear{Kotowski}{Kotowski}{2019}]{kotowski2019perfectly}
\textsc{Kotowski, M.~H.} (2019): \enquote{A Perfectly Robust Approach to
  Multiperiod Matching Problems,} .

\bibitem[\protect\citeauthoryear{Kurino}{Kurino}{2009}]{kurino2009credibility}
\textsc{Kurino, M.} (2009): \enquote{Credibility, efficiency and stability: a
  theory of dynamic matching markets,} \emph{Jena economic research papers,
  JENA}.

\bibitem[\protect\citeauthoryear{Li}{Li}{1993}]{li1993competitive}
\textsc{Li, S.} (1993): \enquote{Competitive matching equilibrium and multiple
  principal-agent models,} Tech. rep., University of Minnesota.

\bibitem[\protect\citeauthoryear{Liu}{Liu}{2023}]{liu2023stability}
\textsc{Liu, C.} (2023): \enquote{Stability in repeated matching markets,}
  \emph{Theoretical Economics}, 18, 1711--1757.

\bibitem[\protect\citeauthoryear{Mauleon, Vannetelbosch, and Vergote}{Mauleon
  et~al.}{2011}]{mauleon2011neumann}
\textsc{Mauleon, A., V.~J. Vannetelbosch, and W.~Vergote} (2011): \enquote{von
  Neumann--Morgenstern farsightedly stable sets in two-sided matching,}
  \emph{Theoretical Economics}, 6, 499--521.

\bibitem[\protect\citeauthoryear{McVitie and Wilson}{McVitie and
  Wilson}{1970}]{mcvitie1970stable}
\textsc{McVitie, D.~G. and L.~B. Wilson} (1970): \enquote{Stable marriage
  assignment for unequal sets,} \emph{BIT Numerical Mathematics}, 10, 295--309.

\bibitem[\protect\citeauthoryear{Nicol{\`o}, Salmaso, and Saulle}{Nicol{\`o}
  et~al.}{2023}]{nicolo2023dynamic}
\textsc{Nicol{\`o}, A., P.~Salmaso, and R.~Saulle} (2023): \enquote{Dynamic
  One-Sided Matching,} Tech. rep., SSRN.

\bibitem[\protect\citeauthoryear{Nicol{\`o}, Salmaso, and Saulle}{Nicol{\`o}
  et~al.}{2024}]{nicolo2024dynamic}
---\hspace{-.1pt}---\hspace{-.1pt}--- (2024): \enquote{Level-$k$ Dynamic
  Matching,} Tech. rep., University of Padua.

\bibitem[\protect\citeauthoryear{Ray and Vohra}{Ray and
  Vohra}{2015}]{ray2015farsighted}
\textsc{Ray, D. and R.~Vohra} (2015): \enquote{The farsighted stable set,}
  \emph{Econometrica}, 83, 977--1011.

\bibitem[\protect\citeauthoryear{Rostek and Yoder}{Rostek and
  Yoder}{2024}]{rostek2024matching}
\textsc{Rostek, M.~J. and N.~Yoder} (2024): \enquote{Matching with Strategic
  Consistency,} \emph{Available at SSRN 2997223}.

\bibitem[\protect\citeauthoryear{Sasaki and Toda}{Sasaki and
  Toda}{1996}]{sasaki1996two}
\textsc{Sasaki, H. and M.~Toda} (1996): \enquote{Two-sided matching problems
  with externalities,} \emph{Journal of Economic Theory}, 70, 93--108.

\end{thebibliography}
\appendix
\section{Omitted statements and proofs from \autoref{sec:consistent-conjectures}}\label{appendix:definitions}
For completeness, in this section, we collect the definitions analogous to those in \autoref{sec:consistent-conjectures} that allow us to describe the construction of the candidate matchings and state the analogue of \autoref{theorem:consistent-conjectures}, \autoref{theorem:cc-app}, solely in terms of the model primitives.
The proof of \autoref{theorem:consistent-conjectures} follows from that of \autoref{theorem:cc-app}.

\paragraph{Non-recursive construction of the candidate matchings} As in the main text, consider the family of conjectures \conjectures. 
\begin{definition}[One-period economy induced by \conjecturet\ at \matchinguptot]\label{definition:one-shot-t}
Given an economy \economyrterminal, a period $t$, and a matching through period $t-1$, \matchinguptot, the one-period economy induced by the collection $\{\conjecturet(\matchinguptot,\agent):\agent\in\calat(\matchinguptot)\cup\calbt(\matchinguptot)\}$ is the length-1 economy defined as follows:
\begin{enumerate}
    \item Agents are $\calat(\matchinguptot)\cup\calbt(\matchinguptot)$,
    \item Utilities $\left\{\payoffact(\agenta,\cdot):\calbt(\matchinguptot)\cup\{\agenta\}\rightarrow\mathbb{R}\vert\agenta\in\calat(\matchinguptot)\right\},\left\{\payoffbct(\cdot,\agentb):\calat(\matchinguptot)\cup\{\agentb\}\rightarrow\mathbb{R}\vert\agentb\in\calbt(\matchinguptot)\right\}$ are given by:
    \begin{enumerate}
\item    For each agent $\agenta\in\calat(\matchinguptot)$, $\payoffact(\agenta,\agenta)=\min\{U_t(\agenta,\matchingb):\matchingb\in \conjecturet(\matchinguptot,\agenta)\}$. Moreover, for $\agentb\in\calbt(\matchinguptot)$, $\payoffact(\agenta,\agentb)=\genericu$.
       \item  For each agent $\agentb\in\calbt(\matchinguptot)$, $\payoffbct(\agentb,\agentb)=\min\{V_t(\agentb,\matchingb):\matchingb\in \conjecturet(\matchinguptot,\agentb)\}$. Moreover, for $\agenta\in\calat(\matchinguptot)$, $\payoffbct\pair=\genericv$.
       \end{enumerate}
\end{enumerate}
We denote by $\economyr_{t,\conjecturet}(\matchinguptot)$ the one-period economy induced by \conjecturet\ at \matchinguptot.
\end{definition}
We can now provide the non-recursive definition of the candidate matchings for \economyrterminal\ given \conjectures:
\begin{align*}
\candidate(\conjectures)=\left\{\matchingstar\in\Matchingsfinal:(\forall t\in\{1,\dots,\terminal\})\;\matchingstar_t|_{\economyr_{t,\conjecturet}(\matchingstaruptot)}\in\stable\left(\economyr_{t,\conjecturet}(\matchingstaruptot)\right)\right\},
\end{align*}
where  (i) in a slight abuse of notation, we use the notation $\candidate(\conjectures)$ to signify the candidate matchings that can be constructed solely as a function of the family of conjectures, \conjectures, for economy \economyrterminal, and (ii) $\matchingstar_t|_{\economyr_{t,\conjecturet}(\matchingstaruptot)}$ is the restriction of $\matchingstar_t$ to the one-period economy $\economyr_{t,\conjecturet}(\matchingstaruptot)$.

\begin{definition}\label{definition:cc-app}
Say \economyrterminal\ has consistent conjectures if for all candidate matchings  $\matchingstar\in\candidate(\conjectures)$ the following holds: 
\begin{align}\label{eq:cc-app}
\text{For all $t\geq1$ and all $\agent$ such that $\matchingstar_t(\agent)=\agent$, we have that }\matchingstar\in\conjecture_t(\matching^{\star^{t-1}},\agent).\end{align}
\end{definition}

The analogous statement to that in \autoref{theorem:consistent-conjectures} now follows:
\begin{theorem}\label{theorem:cc-app}
Fix \economyrterminal\ and suppose it has consistent conjectures. Then, the set of \conjectures-solutions for \economyrterminal\ is nonempty. In particular, $\candidate(\conjectures)\subset\sol_{\terminal}^\conjectures(\economyrterminal)$.
\end{theorem}

Even though the argument leading to the statement of \autoref{theorem:consistent-conjectures} basically provides the proof of that theorem and also of \autoref{theorem:cc-app}, we provide a self-contained proof in the notation in this appendix for completeness:
\begin{proof}[Proof of \autoref{theorem:cc-app}]
Fix a candidate matching $\matchingstar\in\candidate(\conjectures)$ that satisfies \eqref{eq:cc-app}. 
Note that for all $t\in\{1,\dots,\terminal\}$ and all $\agent\in\calat(\matchingstaruptot)\cup\calbt(\matchingstaruptot)$, \matchingstar\ satisfies parts \ref{itm:ira} and \ref{itm:irb} of \autoref{definition:sol-given-c} whenever $\matchingstar_t(\agent)\neq\agent$. Similarly, it satisfies part \ref{itm:ps} of \autoref{definition:sol-given-c} for pairs $\pair\in\calat(\matchinguptot)\times\calbt(\matchinguptot)$ such that $\matchingstar_t(\agenta)\neq\agenta,\matchingstar_t(\agentb)\neq\agentb$.

%
%
%

Consider now a period $t$ and an agent $\agent\in\calat(\matchinguptot)\cup\calbt(\matchinguptot)$ such that $\matchingstar_t(\agent)=\agent$ and note \eqref{eq:cc-app} implies that under \matchingstar, \agent\ is getting at least the payoff of \agent's worst conjectured matching under $\conjecturet(\matchingstaruptot,\agent)$. Given the statement above, \matchingstar\ satisfies parts \ref{itm:ira} and \ref{itm:irb} of \autoref{definition:sol-given-c} for all $\agent\in\calat(\matchinguptot)\cup\calbt(\matchinguptot)$.

Finally, suppose $\matchingstar_t$ fails part \ref{itm:ps} of \autoref{definition:sol-given-c} for some pair \pair\ in $\cala_t(\matching^{\star^{t-1}})\times\calb_t(\matching^{\star^{t-1}})$. Then at least one of the members of the pair must be single under $\matchingstar_t$. Without loss of generality, assume $\matchingstar_t(\agenta)=\agenta$. Then, $u(\agenta,\agentb)>U_t(\agenta,\matchingstar)\geq\min\{U_t(\agenta,\matchingb):\matchingb\in\conjecture_t(\matching^{\star^{t-1}},\agenta)\}$, where the second inequality follows from $\matchingstar\in\conjecture_t(\matching^{\star^{t-1}},\agenta)$. This chain of inequalities contradicts the definition of $\matchingstar_t$. Thus, $\matchingstar\in\sol_{\terminal}^\conjectures(\economyrterminal)$ and the result follows.
\end{proof}

\paragraph{Solution concept based on the conjectures in \autoref{eq:eqbm-cont}} In \autoref{sec:consistent-conjectures}, we argued that the period-1 conjectures defined in \autoref{eq:eqbm-cont} lead to nonempty solutions for \economyrterminal\ \emph{provided}  the solution concept for the continuation economy, $\sol_{\terminal-1}^\conjectures$, is also nonempty. For completeness, we now define the conjectures and the solution concept determined by recursively applying \autoref{eq:eqbm-cont}. This allows us to complete the argument implicit in that discussion. Specifically, a solution concept as defined in \autoref{definition:sol-given-c}, with conjectures consisting of all matchings whose continuations are solutions in the continuation economy, is always nonempty. In what follows, we use the notation \agree\ for the conjectures to distinguish them from those defined in \autoref{eq:eqbm-cont}, and we denote by \solp\ the induced solution concept.

Fix an economy of length \terminal. Recall that \autoref{observation:stability} implies that no matter what the conjectures are, the set of \conjectures-solutions for an economy of length $1$ coincide with the set of stable matchings. Therefore, fix a matching through period $\terminal-2$, $\matching^{\terminal-2}$. For $\agent\in\cala_{\terminal-1}\left(\matching^{\terminal-2}\right)\cup\calb_{\terminal-1}\left(\matching^{\terminal-2}\right)$, define \agent's conjectures as follows:
\begin{align*}
\agree_{\terminal-1}(\matching^{\terminal-2},\agent)=\left\{\matchingb\in\Matchingsfinal(\matching^{\terminal-2},\agent):\matchingb_\terminal|_{\economyrterminal_\terminal(\matchingb^{\terminal-1})}\in\stable\left(\economyrterminal_\terminal(\matchingb^{\terminal-1})\right)\right\}.
\end{align*}
Let $\solp_{\terminal-1}\left(\economyrterminal_{\terminal-1}\left(\matching^{\terminal-2}\right)\right)$ denote the set of $\agree_{\terminal-1}$-solutions for $\economyrterminal_{\terminal-1}\left(\matching^{\terminal-2}\right)$.

Recursively, suppose we have defined $\solp_{\terminal-t}$ for economies of length $\terminal-t$, and fix a matching through period $t$, \matchinguptot. For $\agent\in\calat(\matchinguptot)\cup\calbt(\matchinguptot)$, \agent's conjectures are given by:
\begin{align*}
\agree_{t}(\matchinguptot,\agent)=\left\{\matchingb\in\Matchingsfinal(\matchinguptot,\agent):(\matchingb_s)_{s=t+1}^\terminal|_{\economyrterminal_{t+1}(\matchingb^t)}\in\solp_{\terminal-t}\left(\economyrterminal_{t+1}(\matchingb^t)\right)\right\},
\end{align*}
and let $\solp_{\terminal-t+1}\left(\economyrterminal_t(\matchinguptot)\right)$ denote the set of solutions given the family $\conjectures_{\dagger,\geq t}=\{\agree_s((\matchinguptot,\matchingb_t^{s-1}),\agent):s\geq t, \matchingb_t^{s-1},\agent\in\cala_s(\matchinguptot,\matching_t^{s-1})\cup\calb_s(\matchinguptot,\matching_t^{s-1})\}$.

The following statement is an immediate corollary of \autoref{theorem:consistent-conjectures} and the definition of the conjectures \agree:
\begin{corollary}\label{corollary:agree}
For all $\terminal\in\naturals$,  $\solp_\terminal$ is a nonempty-valued correspondence.
\end{corollary}
\autoref{corollary:agree} then shows existence issues for solution concepts in dynamic matching markets do not arise solely from the perfection requirement---that is, that the continuation matching also abides by the solution concept---but from restrictions the conjectures place on the period-$t$ matchings. 

\autoref{corollary:agree} extends to dynamic matching markets with irreversible matching the existence result for repeated matching in \cite{kotowski2019perfectly}. Indeed, conjectures in \cite{kotowski2019perfectly} consist of all matchings whose continuations are solutions in the continuation economy.

\subsection{Rational expectations and omitted details from \autoref{example:re}}\label{appendix:re}
In this section, we formally define the rational expectations solution, discussed at the end of \autoref{sec:consistent-conjectures}. We then provide supporting details for the derivation in \autoref{example:re}.

\paragraph{Rational expectations solution} The rational expectations solution is defined via a double recursion on the length of the economy and the number of agents available to match in the first period. Below, we proceed in three steps. First, we introduce a piece of notation that simplifies the exposition. Second, we define the rational expectations solution for an economy of length $1$, which gives us the initial step in the recursion on the length of the economy. Third, we define the rational expectations solution for a length-\terminal\ economy.

\underline{\textit{Notation}} Fix a period-$t$ matching \matchingt\ defined over $\arrivalauptot\cup\arrivalbuptot$ and let $\agent\in \arrivalauptot\cup\arrivalbuptot$ be such that $\matchingt(\agent)=\agent$. Define $\matchingt\setminus\agent$ to be the period-$t$ matching with domain and codomain $\arrivalauptot\cup\arrivalbuptot\setminus\{\agent\}$ such that for all $\agent^\prime\neq\agent$, $(\matchingt\setminus\agent)(\agent^\prime)=\matchingt(\agent^\prime)$.

\underline{\textit{Economies of length $\terminal=1$}} We first note for economies of length $1$, the rational expectations solution corresponds to the set of stable matchings (\autoref{definition:galeshapley}). That is, for all $\economyr^1=(\arrivala_1,\arrivalb_1)$, $\re_1(\economyr^1)=\stable(\economyr^1)$.

\underline{\textit{Economies of length $\terminal\geq2$}} Fix an economy \economyrterminal\ and suppose we have defined the rational expectations solution for all economies of length $t\in\{1,\dots,\terminal-1\}$. 

Given any set $\coalition\subseteq\arrivala_1\cup\arrivalb_1$, let $\economyrterminal_{\setminus\coalition}$ denote the length-$\terminal$ economy in which the agents in  \coalition\ arrive in \periodtwo, rather than in \periodone; that is,
\begin{align*}
\economyrterminal_{\setminus\coalition}=(\arrivala_1\setminus\coalition,\arrivalb_1\setminus\coalition,\arrivala_2\cup(\arrivala_1\cap\coalition),\arrivalb_2\cup(\arrivalb_1\cap\coalition),\dots,\arrivala_\terminal,\arrivalb_\terminal).
\end{align*}
Let $N$ denote the cardinality of $\arrivala_1\cup\arrivalb_1$. To define the rational expectations solution for \economyrterminal, we need to define the rational expectations solution for $\economyrterminal_{\setminus\coalition}$ for $\coalition\subseteq\arrivala_1\cup\arrivalb_1$ for \coalition\ of cardinality $1,\dots, N$, starting with coalitions of size $N$: 

\begin{enumerate}[leftmargin=*]
\item When \coalition\ has cardinality $N$, the rational expectations solution of $\economyrterminal_{\setminus\coalition}$, $\re_\terminal(\economyrterminal_{\setminus\coalition})$ corresponds to the rational expectations solution of the induced length $\terminal-1$ economy, $\re_{\terminal-1}(\economyrterminal_{\setminus\coalition})$. Formally, 
\begin{align*}
\re_{\terminal}(\economyrterminal_{\setminus\coalition})=\{\matchingb\in\Matchingsfinal(\economyrterminal_{\setminus\coalition}):(\matchingb_{s})_{s=2}^\terminal\in\re_{\terminal-1}(\economyrterminal_{\setminus\coalition})\}.
\end{align*}

%
\item Suppose \coalition\ has cardinality $m\in\{1,\dots,N-1\}$ and $\re_\terminal$ has been defined for $\economyrterminal_{\setminus\coalitionb}$ for all $\coalitionb:\coalition\subseteq\coalitionb$. For each $\agent\in\arrivala_1\cup\arrivalb_1\setminus\coalition$, define
\begin{align*}
\recbeg(\economyrterminal_{\setminus\coalition},\agent)=\{\matchingb\in\Matchingsfinal(\economyrterminal_{\setminus\coalition},\agent):(\matchingb_1\setminus\agent,(\matchingb_s)_{s=2}^\terminal)\in\re_\terminal(\economyrterminal_{\setminus(\coalition\cup\{\agent\})})\}.
\end{align*}
The rational expectations solution for $\economyrterminal_{\setminus\coalition}$, $\re_\terminal(\economyrterminal_{\setminus\coalition})$, is defined as the subset of $\Matchingsfinal(\economyrterminal_{\setminus\coalition})$ that satisfies \autoref{definition:recursive} when
period-1 conjectures are given by $\{\recbeg(\economyrterminal_{\setminus\coalition},\agent):\agent\in\arrivala_1\cup\arrivalb_1\setminus\coalition\}$ and continuation matchings are elements of $\re_{\terminal-1}$. 
\end{enumerate}

\paragraph{Omitted details from \autoref{example:re}}  In what follows, we provide supporting details for the derivation in \autoref{example:re}. First, note should no one match in period $1$, a unique stable matching exists in \periodtwo, which leads to the matching:
\begin{align*}
\matching_\emptyset=\left(\begin{array}{lcl}&\emptyset&\\\hline
\agenta_1&\_\_&\agentb_2\\
\agenta_2&\_\_&\agentb_4\\
\agenta_3&\_\_&\agentb_3\\
\agenta_4&\_\_&\agentb_1\end{array}\right).
\end{align*}
It follows that $\matching_\emptyset$ is the unique rational expectations solution in any two-period economy in which at least $\agentb_1,\agentb_2$ are part of the \periodtwo-arrivals.

Second, consider $\economyr_{\setminus\{\agentb_1\}}^2$. We argue $\matching_{\agentb_1}$ is a rational expectations solution. To illustrate, note $\agentb_2$ does not wish to block with $\agenta_1$, and $\agenta_3$ cannot do better than by matching in \periodone\ with $\agentb_2$. The only possibility is that $\agentb_2$ waits to be matched. However, $\matching_\emptyset$ is a valid conjecture for $\agentb_2$ in  $\economyr_{\setminus\{\agentb_1,\agentb_2\}}^2$, and $\agentb_2$ prefers $\agenta_3$ to matching with $\agenta_1$ in \periodtwo.

Third, consider  $\economyr_{\setminus\{\agenta_3,\agentb_1\}}^2$. We show $\agentb_1$ is always matched to $\agenta_4$ in any rational expectations solution, and hence, $\agentb_1$ cannot match with $\agenta_1$ in a rational expectations solution for  $\economyr_{\setminus\{\agenta_3\}}^2$. In  $\economyr_{\setminus\{\agenta_3,\agentb_1\}}^2$, three possibilities exist for a matching that is a rational expectations solution: (i) no one matches in \periodone, which would lead to $\matching_\emptyset$ (though note $\matching_\emptyset$ is blocked by $(\agenta_1,\agentb_2)$ in  $\economyr_{\setminus\{\agenta_3,\agentb_1\}}^2$), (ii) $\agenta_1$ matches with $\agentb_2$, and (iii) $\agenta_2$ matches with $\agentb_2$. Under (ii), note the outcome of side-\arrivala\ proposing deferred acceptance in \periodtwo\ among $\agenta_2,\agenta_3,\agenta_4,\agentb_1,\agentb_3,\agentb_4$ is
\begin{align*}
\left.\begin{array}{lcl}\agenta_2&\_\_&\agentb_4\\
\agenta_3&\_\_&\agentb_3\\
\agenta_4&\_\_&\agentb_1\end{array}\right.,
\end{align*}
so any rational expectations solution for  $\economyr_{\setminus\{\agenta_3,\agentb_1\}}^2$ that matches $\agenta_1,\agentb_2$ in \periodone\ (if any) gives $\agentb_1$ at least the payoff of matching with $\agenta_4$. Similarly, under (iii), note the outcome of side-\arrivala\ proposing deferred acceptance in \periodtwo\ among $\agenta_1,\agenta_3,\agenta_4,\agentb_1,\agentb_3,\agentb_4$ is
\begin{align*}
\left.\begin{array}{lcl}\agenta_1&\_\_&\agentb_4\\
\agenta_3&\_\_&\agentb_3\\
\agenta_4&\_\_&\agentb_1\end{array}\right.,
\end{align*}
so that any rational expectations solution for  $\economyr_{\setminus\{\agenta_3,\agentb_1\}}^2$ that matches $\agenta_2,\agentb_2$ in \periodone\ (if any) gives $\agentb_1$ at least the payoff of matching with $\agenta_4$. It follows that in $\economyr_{\setminus\{\agenta_3\}}^2$, $\agentb_1$ has to match with $\agenta_4$.

Fourth, consider $\economyr_{\setminus\{\agenta_2\}}^2$. We argue  $\matching_{\agenta_2}$ is a rational expectations solution for that economy. To do so, we need to argue $\agentb_1$ cannot block. Consider then the economy $\economyr_{\setminus\{\agenta_2,\agentb_1\}}^2$ and note $\matching_{\agentb_1}$ is a rational expectations solution for the same reason it is a rational expectations solution in $\economyr_{\setminus\{\agentb_1\}}^2$. Thus, by blocking $\matching_{\agenta_2}$, $\agentb_1$ cannot expect to get more than the payoff of matching with $\agenta_1$ in \periodtwo. It follows that $\matching_{\agenta_2}$ is a rational expectations solution for $\economyr_{\setminus\{\agenta_2\}}^2$.

Finally, we argue \matchingstar\ in the example is a candidate matching as defined in \autoref{sec:consistent-conjectures}. Note that because of $\matching_{\agenta_2}$, $\agenta_2$ is willing to match with $\agentb_2$ in the one-period economy induced by the conjectures, and similarly, $\agentb_1$ is willing to match with $\agenta_1$. Because $\agenta_2$ and $\agentb_2$ top-rank each other in the one-period economy induced by the conjectures, they match together in any stable matching, which then justifies $\agenta_1$ and $\agentb_1$ matching, and $\agenta_3$ remaining unmatched. When $\{\agenta_3,\agenta_4\}$ are the \periodtwo\ agents on side \arrivala, a unique stable matching exists, which is specified by $\matchingstar_2$.

\section{Omitted proofs from \autoref{sec:solutions}}\label{appendix:solutions}
\begin{proof}[Proof of \autoref{proposition:cvr}]
The proof of \autoref{proposition:cvr} proceeds by induction on the economy length, $\terminal\in\naturals$. For $\terminal=1$, the correspondence $\dsh_1$ is nonempty valued (cf. \autoref{observation:stability}) and the statement about $\conjecture_{\dsh,1}$ is vacuous.

Assume then we know the correspondence $\dsh_t$ is nonempty valued for $t\in\{1,\dots,\terminal-1\}$. As is clear from the main text, to establish the main result, showing that for any \economyrterminal, the period-1 conjectures $\{\conjecture_{\dsh,1}(\cdot,\agent):\agent\in\arrivala_1\cup\arrivalb_1\}$ are nonempty and satisfy consistency is enough.

Fix an economy \economyrterminal. For each $\agent\in\arrivala_1\cup\arrivalb_1$, let $\conjecture_0(\emptyset,\agent)=\Matchingsfinal(\emptyset,\agent)$ and recursively define for $n\geq1$
\begin{align*}
\conjecture_n(\emptyset,\agent)=\left\{\matchingb\in\Matchingsfinal(\emptyset,\agent):\begin{array}{l}
\matchingb_1\in\stable(\economyr_{1,\conjecture_{n-1}}^{\matchingb_1}),
(\matchingb_t)_{t=2}^\terminal|_{\economyr_2^\terminal(\matchingb_1)}\in\dsh_{T-1}(\economyr_2^\terminal(\matchingb_1))
\end{array}\right\}.
\end{align*}
We observe that for all $\agent\in\arrivala_1\cup\arrivalb_1$, the following hold. First,  $\conjecture_{n+1}(\emptyset,\agent)\subseteq\conjecture_{n}(\emptyset,\agent)$ because in each step, we are increasing the continuation values of period-1 agents other than \agent. Formally, by definition, we have that for all $\agent\in\arrivala_1\cup\arrivalb_1$, $\conjecture_1(\emptyset,\agent)\subseteq\conjecture_0(\emptyset,\agent)$. Suppose we have shown $\conjecture_n(\emptyset,\agent^\prime)\subseteq\conjecture_{n-1}(\emptyset,\agent^\prime)$ for all $\agent^\prime\in\arrivala_1\cup\arrivalb_1$; we proceed to show  $\conjecture_{n+1}\subseteq\conjecture_n$. Fix $\agent\in\arrivala_1\cup\arrivalb_1$ and $\matchingb\in\conjecture_{n+1}(\emptyset,\agent)$. We only need to show that if $\matchingb_1\in\stable(\economyr_{1,\conjecture_{n}}^{\matchingb_1})$, $\matchingb_1\in\stable(\economyr_{1,\conjecture_{n-1}}^{\matchingb_1})$. To this end, let $\agent^\prime$ be such that $\matchingb_1(\agent^\prime)\neq\agent^\prime$. By construction, $\agent^\prime$ prefers $\matchingb$ to the worst conjectured matching in $\conjecture_n(\emptyset,\agent^\prime)\subseteq\conjecture_{n-1}(\emptyset,\agent^\prime)$. Thus, $\agent^\prime$ also prefers $\matchingb$ to the worst conjectured matching in $\conjecture_{n-1}(\emptyset,\agent^\prime)$. Thus, $\matchingb\in\conjecture_{n}(\emptyset,\agent)$. Second, because everything is finite, $\conjecture_\infty(\emptyset,\agent)=\lim_{n\rightarrow\infty}\conjecture_n(\emptyset,\agent)$ is well-defined. Third,  because $\{\matchingb\in\Matchingsfinal(\emptyset,\agent):(\matchingbt)_{t=2}^\terminal\in\dsh_{\terminal-1}\left(\economyr_2^\terminal(\matchingb_1)\right)\}\subseteq\conjecture_n(\emptyset,\agent)$ for all $n$ and is nonempty, $\conjecture_{\infty}(\emptyset,\agent)$ is nonempty.

We claim 
\begin{align}\label{eq:identity}
\conjecture_\infty(\emptyset,\agent)=\left\{\matchingb\in\Matchingsfinal(\emptyset,\agent):\begin{array}{l}
\matchingb_1\in\stable(\economyr_{1,\conjecture_{\infty}}^{\matchingb_1}),
(\matchingb_s)_{s=2}^\terminal|_{\economyr_2^\terminal(\matchingb_1)}\in\dsh_{\terminal-1}(\economyr_2^\terminal(\matchingb_1))
\end{array}\right\},
\end{align}
and hence $\conjecture_\infty=\conjecture_{\dsh,1}$.

To see \autoref{eq:identity} holds, note the following:
\begin{enumerate}[leftmargin=*]
\item If $\overline{N}$ exists such that for all $\agent\in \arrivala_1\cup \arrivalb_1$, $\conjecture_{\overline{N}}(\emptyset,\agent)=\conjecture_{\overline{N}+1}(\emptyset,\agent)$, then for all $N\geq\overline{N}$ and all $\agent\in\arrivala_1\cup\arrivalb_1$, it follows that $\conjecture_N(\emptyset,\agent)=\conjecture_{\overline{N}+1}(\emptyset,\agent)$. This follows from the definition of $\conjecture_{n}$.

\item Because the sets $\conjecture_n(\emptyset,\agent)$ are finite, as is the set of period-1 agents,  $\overline{N}<\infty$ exists such that for all $\agent\in\arrivala_1\cup\arrivalb_1$, $\conjecture_{\overline{N}}=\conjecture_{\overline{N}+1}$. Toward a contradiction, suppose that for all $N\in\mathbb{N}$  $\agent_N\in\arrivala_1\cup\arrivalb_1$ exists such that $\conjecture_N(\emptyset,\agent_N)\neq \conjecture_{N+1}(\emptyset,\agent_{N})$. Because the set $A_1\cup B_1$ is finite, we can find an agent \agent\ and a subsequence $(n^\agent)$ such that $\conjecture_{n^\agent}(\emptyset,\agent)\neq \conjecture_{n^\agent+1}(\emptyset,\agent)$. Thus, $\conjecture_{n^\agent+1}(\emptyset,\agent)\subsetneq \conjecture_{n^\agent}(\emptyset,\agent)$. Since $\conjecture_{n^\agent}(\emptyset,\agent)$ is finite, this means that eventually, $\conjecture_{n^\agent}(\emptyset,\agent)=\emptyset$, a contradiction. Hereafter, let $\overline{N}$ denote the smallest $n$ such that for all $\agent\in\arrivala_1\cup\arrivalb_1$, $\conjecture_{\overline{N}}(\emptyset,\agent)=\conjecture_{\overline{N}+1}(\emptyset,\agent)\equiv \conjecture_\infty(\emptyset,\agent)$.

\item Suppose $\agent\in\arrivala_1\cup\arrivalb_1$ exists such that $\matchingb\in \conjecture_\infty(\emptyset,\agent)$, but $\matchingb$ is not an element of the set on the right-hand side of \autoref{eq:identity}. Then, without loss of generality, $\agenta\in\arrivala_1\cup\arrivalb_1\setminus\{\agent\}$ exists such that $\matchingb_1(\agenta)\neq\agenta$ and
$
\min_{\matchingc\in \conjecture_\infty(\emptyset,\agenta)}U_1(\agenta,\matchingc)>u(a,\matchingb_1(\agenta)).
$ 
This implies $\matchingb\notin \conjecture_{\overline{N}+2}(\emptyset,\agent)$, contradicting that $\conjecture_{\overline{N}+1}(\emptyset,\agent)=\conjecture_\infty(\emptyset,\agent)$. 
\end{enumerate}
We conclude that for all $\agent\in\arrivala_1\cup\arrivalb_1$, \autoref{eq:identity} holds, and hence, $\conjecture_\infty=\conjecture_{\dsh,1}$. Thus, 
for all $\agent\in\arrivala_1\cup\arrivalb_1$, $\solonec(\emptyset,\agent)\neq\emptyset$. 

We now show $\conjecture_{\dsh,1}$ satisfy \ref{eq:cc}. To this end, construct a candidate \matchingstar\ as described in \autoref{sec:consistent-conjectures} using $\conjecture_{\dsh,1}$ and the correspondence $\dsh_{\terminal-1}$, which by the inductive hypothesis is nonempty. Fix a period-1 agent, $\agent\in\arrivala_1\cup\arrivalb_1$ such that $\matchingstar_1(\agent)=\agent$. Note \matchingstar\ satisfies the following: First, the continuation matching $(\matchingstar_t)_{t=2}^\terminal\in\dsh_{\terminal-1}(\economyrterminal_2(\matchingstar_1))$. Second, for all $\agent^\prime\in\arrivala_1\cup\arrivalb_1$ such that $\matchingstar_1(\agent^\prime)\neq\agent^\prime$, we have that $\agent^\prime$ is matched to someone they prefer to the worst matching in $\conjecture_{\dsh,1}(\emptyset,\agent^\prime)$. Finally, note $\matchingstar_1$ has no blocking pairs \pair\ such that $\matchingstar_1(\agenta)\neq\agenta,\matchingstar_1(\agentb)\neq\agentb$. Thus, $\matchingstar_1\in\stable(\economyr_{1,\solonec}^{\matchingstar_1})$. Hence, $\matchingstar\in\conjecture_{\dsh,1}(\emptyset,\agent)$ so that \ref{eq:cc} holds. 

\autoref{theorem:consistent-conjectures} then implies $\matchingstar\in\dsh_\terminal(\economyrterminal)$. Because the proof only used that \matchingstar\ is \emph{a} candidate matching, this completes the inductive step.
\end{proof}

\begin{proof}[Proof of \autoref{proposition:sds-refines}]
The proof proceeds by complete double induction on the length of the economy, \terminal, and the number of agents arriving in period 1, $n$. Let $\mathrm{P}(\terminal,n)$ denote the following inductive statement:

$\mathrm{P}(\terminal,n)$: For all economies of length \terminal\ with $n$ agents in period $1$, $\sds_\terminal\subseteq\ds_\terminal$. 

That $\mathrm{P}(1,n)=1$  for all $n\in\naturals$ follows from noting that for all length-1 economies, $\economyr_1$, $\sds_1(\economyr_1)=\ds_1(\economyr_1)=\stable_1(\economyr_1)$. Similarly, $\mathrm{P}(2,1)=1$ because in a length-2 economy \economyrtwo\ in which only one agent arrives in \periodone, the only candidates for (sophisticated) dynamically stable matchings are matchings $(\matching_1,\matching_2)$ whose continuation matching is stable for the length-1 economy $(A_1\cup A_2,B_1\cup B_2)$. That is, for any length-2 economy in which only one agent arrives in period 1, the sets of dynamically stable and sophisticated dynamically stable matchings coincide.

The rest of the proof consists of the following two inductive steps. First, if $\terminal\geq 1$ and $n\geq 1$ exist such that, for all $n^\prime\leq n$, $\mathrm{P}(\terminal,n^\prime)=1$, then $\mathrm{P}(\terminal,n+1)=1$. This first step then proves that $\mathrm{P}(\terminal,\cdot)=1$. Second, if  \terminal\ exists such that, for all $\terminal^\prime <\terminal$, $\mathrm{P}(T^\prime,\cdot)=1$, $\mathrm{P}(\terminal+1,1)=1$. Together, these steps prove that the inductive statement is true for all $\terminal\in\naturals$ and $n\in\naturals$.\footnote{Indeed, $\mathrm{P}(1,\cdot)=\mathrm{P}(2,1)=1$ by our initial steps. Moreover, $\mathrm{P}(2,1)=1$ and the first inductive step shows that $\mathrm{P}(2,\cdot)=1$. The second inductive step then proves $\mathrm{P}(3,1)=1$. Applying the first and second inductive steps sequentially shows $\mathrm{P}(3,\cdot)=1$, and hence, $\mathrm{P}(4,1)=1$, respectively. Iterating this process shows $P(\terminal,n)=1$ for all \terminal\ and $n$.}

%
%

Fix $\terminal\geq 2$ and $n\geq1$. Fix an economy of length \terminal\ with $n$ agents in \periodone, and denote it by \economyrterminal. By the inductive hypothesis, suppose the following two properties hold. First, for all $\terminalb<\terminal$ and all $n^\prime\in\naturals$, $\sds_{\terminalb}(\economyr^{\terminalb})\subseteq\ds_{\terminalb}(\economyr^{\terminalb})$ for all economies with $n^\prime$ agents in \periodone. By the discussion above, this implies $P(\terminal,1)=1$, and hence, we can take $n\geq2$ without loss of generality. Second, for all economies of length \terminal\ with $n^\prime< n$ agents in \periodone, $\economyr^{\prime^\terminal}$, $\sds_\terminal(\economyr^{\prime^\terminal})\subseteq\ds_\terminal(\economyr^{\prime^\terminal})$. We show  $\sds_\terminal(\economyrterminal)\subseteq\ds_\terminal(\economyrterminal)$. 

It suffices to show that for all $\agent\in\arrivala_1\cup\arrivalb_1$, $\conjecture_{\sds,1}(\economyrterminal,\agent)\subseteq\conjecture_{\ds,1}(\economyrterminal,\agent)$, 
where we now make apparent that the conjectures in dynamic stability depend on the economy. To do so, we show that for all $l\in\naturals_0$,  $\addon_l(\economyrterminal,\agent)\subseteq\conjecture_{\ds,1}(\economyrterminal,\agent)$.

Consider first $l=0$. Recall that
\begin{align}\label{eq:incl-0}
\addon_0(\economyrterminal,\agent)&=\left\{\matchingb\in\Matchingsfinal(\emptyset,\agent;\economyrterminal):(\matchingb_1\setminus\agent,(\matchingb_s)_{s=2}^\terminal)\in\sds_\terminal(\economyrterminal_{\setminus\left\{\agent\right\}})\right\}\nonumber\\
&\subseteq\left\{\matchingb\in\Matchingsfinal(\emptyset,\agent;\economyrterminal):(\matchingb_1\setminus\agent,(\matchingb_s)_{s=2}^\terminal)\in\ds_\terminal(\economyrterminal_{\setminus\left\{\agent\right\}})\right\}\nonumber\\
&\subseteq\conjecture_{\ds,1}(\economyrterminal,\agent),
\end{align}
where the first inclusion follows by assumption, and the second inclusion, from observing that dynamic stability imposes more conditions than the conjectures of agent \agent\ under dynamic stability on the period-$1$ matching.

The above inclusion implies 
\begin{align}\label{eq:candidate-incl}
\candidate(\emptyset,\addon_0,\sds_{\terminal-1})&=\{\matchingb\in\Matchingsfinal(\economyrterminal):\text{ (i) } \matchingb_1\in\stable(\economyr_{1,\addon_0})\text{ (ii) }(\matchingb_s)_{s=2}^\terminal|_{\economyrterminal_2(\matchingb_1)}\in\sds_{\terminal-1}(\economyrterminal_2(\matchingb_1))\}\nonumber\\
&\subseteq\{\matchingb\in\Matchingsfinal(\economyrterminal):\text{ (i) } \matchingb_1\in\stable(\economyr_{1,\conjecture^{DS}})\text{ (ii) }(\matchingb_s)_{s=2}^\terminal|_{\economyrterminal_2(\matchingb_1)}\in\ds_{\terminal-1}(\economyrterminal_2(\matchingb_1))\}\nonumber\\
&\subseteq\ds_\terminal(\economyrterminal),
\end{align}
where the first inclusion follows from $F_0(\economyrterminal,\agent)\subseteq\conjecture_{\ds,1}(\economyrterminal,\agent)$ for all $\agent\in\arrivala_1\cup\arrivalb_1$ and $\sds_{\terminal-1}\subseteq\ds_{\terminal-1}$, and the second inclusion follows from noting the second line describes the candidate matchings for dynamic stability. 
Thus,
\begin{align*}
\addon_1(\economyrterminal,\agent)=\addon_0(\economyrterminal,\agent)\cup\{\matchingb\in\candidate(\emptyset,\addon_0,\sds_{\terminal-1}):\matchingb_1(\agent)=\agent\}\subseteq\conjecture_{\ds,1}(\economyrterminal,\agent),
\end{align*}
where the inclusion follows from Equations \ref{eq:incl-0} and \ref{eq:candidate-incl}. Repeating the argument in \autoref{eq:candidate-incl} for $l\geq1$ delivers that $\conjecture_{\sds,1}(\economyrterminal,\agent)\subseteq\conjecture_{\ds,1}(\economyrterminal,\agent)$ and hence that $\sds_\terminal(\economyrterminal)\subseteq\ds_\terminal(\economyrterminal)$. 
\end{proof}

\paragraph{Omitted details from \autoref{example:cvr-not-sds}} In what follows, we provide supporting details for the derivation in \autoref{example:cvr-not-sds}. We recursively define the set of sophisticated dynamically stable matchings for the economy in the example, starting from the case in which all \periodone\ agents wait to be matched. In this example, sophisticated dynamic stability coincides with the rational expectations solution; in particular, the recursion to build the conjectures \sdsc, described in the main text, is not necessary.

Note that if nobody matches in \periodone, a unique stable matching exists in \periodtwo, which leads to the matching $\matching_\emptyset$ in the main text. It follows that $\matching_\emptyset$ is the unique sophisticated dynamically stable matching in any two-period economy in which $\agentb_1$ is part of the \periodtwo-arrivals. 

Consider now the case in which two agents in $\arrivala_1$ wait to be matched, so that the induced economies are $\economyrtwo_{\setminus\{\agenta_i,\agenta_j\}}$ for $i,j\in\{1,2,3\}$, $i\neq j$. Using the fact that in each of these economies, the conjectures of a period-1 agent who waits to be matched coincides with $\matching_\emptyset$, we conclude the following:
\begin{align}\label{eq:sds-2}
\sds_{2}(\economyrtwo_{\setminus\{\agenta_2,\agenta_3\}})&=\{\matching^R,\matching^C\},
&\sds_{2}(\economyrtwo_{\setminus\{\agenta_1,\agenta_3\}})&=\{\matching^L\},
&\sds_{2}(\economyrtwo_{\setminus\{\agenta_1,\agenta_2\}})&=\{\matching_\emptyset\}.
\end{align}
In $\economyrtwo_{\setminus\{\agenta_2,\agenta_3\}}$, $\agenta_1$ and $\agentb_1$ match with each other in \periodtwo\ if either of them waits to be matched. Because they both discount the future, neither of them blocks $\matching^R$ or $\matching^C$. Instead, $\matching_\emptyset$ is not sophisticated dynamically stable, because $\agenta_1,\agentb_1$ block by matching together in \periodone. Similarly, in $\economyrtwo_{\setminus\{\agenta_1,\agenta_3\}}$, agent $\agenta_2$ expects to match with $\agentb_3$ if they wait to be matched. Both $\agenta_2,\agentb_1$ prefer to match together to waiting to be matched, so that $\matching^L$ is the unique sophisticated dynamically stable matching in that economy. Finally, in $\economyrtwo_{\setminus\{\agenta_1,\agenta_2\}}$, $\agenta_3$ expects to match with $\agentb_2$ if they wait to be matched, and this matching is preferred to matching with agent $\agentb_1$. Thus, $\matching_\emptyset$ is the unique sophisticated dynamically stable matching. 

Consider now the case in which one agent in $\arrivala_1$ waits to be matched, so that the induced economies are $\economyrtwo_{\setminus\{\agenta_i\}}$ for $i\in\{1,2,3\}$. Using the sophisticated dynamically stable matchings in \autoref{eq:sds-2}, we can easily show 
\begin{align}\label{eq:sds-1}
\sds_2(\economyrtwo_{\setminus\{\agenta_1\}})&=\{\matching^L\},&\sds_2(\economyrtwo_{\setminus\{\agenta_2\}}) &=\{\matching^C,\matching^R\},&\sds_2(\economyrtwo_{\setminus\{\agenta_3\}})&=\{\matching^L\}.
\end{align}
We explain the intuition behind $\sds_2(\economyrtwo_{\setminus\{\agenta_2\}})$, because it is used to show why $\matching^L$ is the unique element of $\sds_2(\economyrtwo)$; similar intuition applies for the others. When $\agenta_2$ is treated as a period-2 agent, the induced economies by either $\agenta_1$ or $\agenta_3$ waiting to be matched are $\economyrtwo_{\setminus\{\agenta_1,\agenta_2\}}$ and  $\economyrtwo_{\setminus\{\agenta_2,\agenta_3\}}$, respectively. Thus, $\agenta_1$ can guarantee at most the payoff from matching with $\agentb_1$ in \periodtwo, whereas $\agenta_3$ can guarantee the payoff from matching with $\agentb_2$ in \periodtwo. It follows that $\agenta_3$ must match in \periodtwo, and $\agenta_1$ must match with $\agentb_1$ in \periodone. 

The above arguments imply  $\agenta_2$ can guarantee at most the payoff from matching with $\agentb_3$ in \periodtwo\ when the economy is \economyrtwo. This justifies that $\agenta_2$ does not block $\matching^L$ by waiting, and thus, it is an element of $\sds_2(\economyrtwo)$. Because $\agenta_1$ and $\agenta_3$ can guarantee the payoffs from matching with $\agentb_2$ and $\agentb_3$, respectively, it follows that $\matching^L$ is the unique sophisticated dynamically stable matching in \economyrtwo.

\end{document}